\documentclass[lettersize,journal]{IEEEtran}
\usepackage{amsmath,amsfonts,algpseudocode} 
\usepackage{algorithm}
\usepackage{array}
\usepackage[caption=false,font=normalsize,labelfont=sf,textfont=sf]{subfig}
\usepackage{textcomp}
\usepackage{stfloats}
\usepackage{url}
\usepackage{verbatim}
\usepackage{graphicx}
\usepackage{cite}
\usepackage{orcidlink}
\usepackage[most]{tcolorbox} 
\usepackage{indentfirst}
\usepackage{flushend}
\usepackage{amsthm} 
\usepackage{bm}
\usepackage{amssymb}
\usepackage{color}
\usepackage{tabularx}
\usepackage{booktabs}
\usepackage{multicol}
\usepackage{multirow}
\usepackage{caption}
\usepackage{float}
\usepackage{diagbox}
\usepackage{tikz}
\usepackage{pifont}
\usepackage{hyperref}
\usepackage{xcolor}
\definecolor{faded}{gray}{0.5}
\definecolor{orange}{RGB}{200,0,100}

\newcommand{\tb}[1]{\textbf{#1}}
\newcommand{\ti}[1]{\textit{#1}}

\newcommand{\norm}[1]{\left\|#1\right\|}

\def \h{ \mathbf h }
\def \k{ \mathbf k }

\def \u{ \mathbf u }
\def \v{ \mathbf v }

\def \x{ \mathbf x }
\def \y{ \mathbf y }

\def \hh{ \hat \h }
\def \hx{ \hat \x }
\def \hy{ \hat \y }

\def \A{ \mathbf A }
\def \B{ \mathbf B }
\def \C{ \mathbf C }
\def \D{ \mathbf D }

\def \H{ \mathbf H }
\def \I{ \mathbf I }
\def \K{ \mathbf K }
\def \I{ \mathbf I }
\def \P{ \mathbf P }
\def \Q{ \mathbf Q }
\def \S{ \mathbf S }
\def \Y{ \mathbf Y }
\def \hC{ \hat \C }
\def \hH{ \hat \H }
\def \bbC{ \mathbb C }
\def \bbE{ \mathbb E }
\def \bbN{ \mathbb N }
\def \bbR{ \mathbb R }

\def \bSigma{ \boldsymbol \Sigma }
\def \bPhi{ \boldsymbol \Phi }
\def \b0{ \boldsymbol 0 }

\def \bpsi{ \boldsymbol \psi } 

\def \BigO{ \mathcal O }
\def \vec{ \operatorname{vec} } 
\DeclareMathOperator*{\argmax}{arg\,max}
\DeclareMathOperator*{\argmin}{arg\,min}
\def \Re{ \text{Re} }
\def \Im{ \text{Im} }

\def \LoS{ \text{LoS} }
\def \NLoS{ \text{NLoS} }

\newtheorem{proposition}{Proposition}

\newtheorem{remark}{Remark}

\newenvironment{italicproof}
    {\itshape}
    {}

\begin{document}
\title{Hybrid Data-Driven SSM for Interpretable and Label-Free mmWave Channel Prediction}

\author{
  Yiyong Sun\raisebox{1ex}{\orcidlink{0009-0006-4823-0202}},
  Jiajun He\raisebox{1ex}{\orcidlink{0000-0003-4304-7354}}, 
  Zhidi Lin\raisebox{1ex}{\orcidlink{0000-0002-6673-511X}}, 
  Wenqiang Pu\raisebox{1ex}{\orcidlink{0000-0003-3923-056X}},~\IEEEmembership{Member,~IEEE}, \\
  Feng Yin\raisebox{1ex}{\orcidlink{0000-0001-5754-9246}},~\IEEEmembership{Senior Member,~IEEE},
  Hing Cheung So\raisebox{1ex}{\orcidlink{0000-0001-8396-7898}},~\IEEEmembership{Fellow,~IEEE}

  \thanks{Y. Sun and F. Yin are with School of Science and Engineering, The Chinese University of Hong Kong, Shenzhen,
  Guangdong, 518172, P. R. China (email: yiyongsun@link.cuhk.edu.cn, yinfeng@cuhk.edu.cn). }
  \thanks{J. He is with Centre for Wireless Innovation (CWI), Queen’s University Belfast, BT3 9DT Belfast, U.K. (e-mail: j.he@qub.ac.uk). }
  \thanks{Z. Lin is with the Department of Statistics and Data Science, National University of Singapore, 117546, Singapore (email: zhidilin@nus.edu.sg).}
  \thanks{W. Pu is with Shenzhen Research Institute of Big Data, Shenzhen,
  Guangdong, 518172, P. R. China (email: wenqiangpu@cuhk.edu.cn). }
  \thanks{H. C. So is with the Department of Electrical Engineering, City University of Hong Kong, Hong Kong, P. R. China (e-mail: hcso@ee.cityu.edu.hk). }
  \thanks{Corresponding author: Feng Yin.}
}

\markboth{Journal of \LaTeX\ Class Files,~Vol.~18, No.~9, September~2020}
{Hybrid Data-Driven SSM for Interpretable and Label-Free mmWave Channel Prediction}

\maketitle

\begin{abstract} 
  Accurate prediction of mmWave time-varying channels is essential for mitigating the issue of \ti{channel aging} in complex scenarios owing to high user mobility. Existing channel prediction methods have limitations: classical model-based methods often struggle to track highly nonlinear channel dynamics due to limited expert knowledge, while emerging data-driven methods typically require substantial labeled data for effective training and often lack interpretability. To address these issues, this paper proposes a novel hybrid method that integrates a data-driven neural network into a conventional model-based workflow based on a state-space model (SSM), implicitly tracking complex channel dynamics from data without requiring precise expert knowledge. Additionally, a novel unsupervised learning strategy is developed to train the embedded neural network solely with unlabeled data. Theoretical analyses and ablation studies are conducted to interpret the enhanced benefits gained from the hybrid integration. Numerical simulations based on the 3GPP mmWave channel model corroborate the superior prediction accuracy of the proposed method, compared to state-of-the-art methods that are either purely model-based or data-driven. Furthermore, extensive experiments validate its robustness against various challenging factors, including among others severe channel variations and high noise levels.
\end{abstract}


\section{Introduction}
\label{sec-intro}
\IEEEPARstart{M}{assive} multiple-input multiple-output (MIMO) systems with millimeter-wave (mmWave) play a crucial role in fifth- and sixth-generation (5G/6G)  wireless networks operating in complicated environments such as vehicle-to-everything (V2X) \cite{tmc_2023_v2x}, intelligent factories \cite{tmc2021delay}, and air-to-ground communication \cite{tmc_2022_air}. These applications necessitate not only high throughput through effective beamforming \cite{tmc_2022_beamforming,tmc_2020_beamforming} but also ubiquitous sensing capabilities for precise localization \cite{gao2023metaloc, he2024cramer} and environment perception \cite{tmc2024sensing,he2023framework}. To enable such emerging and smart applications, accurate and agile estimation of instantaneous channel state information (CSI) is in high demand\cite{tmc2021robust, tmc2024efficient,jiang2019exploiting}.

Conventional CSI estimation methods involve periodically sending pilot signals to mobile user equipment (UE) \cite{tmc2020beam,tmc2021context,tmc2023online}, leading to challenges in complex scenarios with rapidly moving UE \cite{he2023modeling}. Specifically, the massive antenna array at the base station (BS) results in unaffordable data processing delay, which in turn leads to inaccurate CSI estimates. The inaccuracies arise from the misalignment between the time at which the CSI is estimated and the time it is used, commonly referred to as the channel aging (or channel outdating) \cite{chopra2021data}. Such challenges can be further exacerbated by the increased user mobility, i.e., the pronounced Doppler effects and fast-changing channels lead to outdated CSI estimation. One may seek more frequent transmission of the pilot signals to facilitate real-time CSI tracking. Nonetheless, this will introduce notable communication service delay and overhead \cite{jiang2019exploiting}.

\subsection{Related Works}
\label{sec-intro-related}
\noindent A promising solution to address rapid channel aging is to predict future CSI directly \cite{baddour2005autoregressive, kashyap2017performance,zhu2019adaptive}. This can be achieved by leveraging its temporal correlation with historical channels without sacrificing additional communication overhead \cite{kim2020massive}. Channel prediction is usually formulated as a time series prediction task \cite{kim2020massive, wei2022channel, jiang2022accurate}. The main challenge lies in accurately capturing the complex temporal pattern (or dynamics) of time-varying channels \cite{jiang2022accurate}. In complex scenarios with high user velocity, there is limited expert knowledge available to explicitly characterize the underlying dynamic process due to the resulting high nonlinearity. Practical solution has been further complicated by the available data, which comprises only a sequence of noisy signals of limited length and without high-quality manual labeling. Existing techniques mainly focus on approximating the channel dynamics and can be broadly categorized into model-based and data-driven methods.

Model-based methods assume that the channel dynamics follow certain analytical functions, such as linear extrapolation \cite{yin2020addressing}, sum-of-sinusoids \cite{wong2005joint}, autoregressive (AR) models \cite{baddour2005autoregressive}, and autoregressive moving average (ARMA) models \cite{kashyap2017performance}. Their limited expressiveness hinders these simplified approximations from accurately matching the actual channel dynamics \cite{jiang2022accurate}, leading to unstable performance under high user mobility. Building on the approximated channel dynamics, recent works in \cite{chopra2021data, kashyap2017performance, kim2020massive} further integrates the signal model to construct a linear SSM, referred to as the model-based SSM, which facilitates a filter-then-predict (FTP) workflow based on the well-established Kalman filtering (KF) \cite{kalman1960new}. Specifically, in this model-based FTP workflow, KF is recursively implemented for error correction before employing the approximated dynamics for prediction. By incorporating physical knowledge of the SSM in a principled manner (i.e., according to the theoretical foundation of KF), the model-based FTP workflow eliminates the requirement for well-labeled CSI data.\footnote{Unlabeled data refers to the received noisy signals. Labeled data is the estimate of the CSI obtained from unlabeled data through signal processing techniques like linear minimum mean squared error (LMMSE) \cite{tse2005fundamentals}.} Nonetheless, the effectiveness of KF demands precise knowledge of the actual channel dynamics to compute the Kalman gain, a crucial weighting matrix in KF. Approximating the highly nonlinear channel dynamics with a linear SSM may cause notable model mismatch, resulting in significantly deviated Kalman gain and, thereby, degraded filtering performance \cite{revach2022kalmannet}. Furthermore, the original objective of the Kalman gain is to minimize filtering errors \cite{sarkka2023bayesian}, which serves as an intermediate step before minimizing prediction errors in the channel prediction task. The model-based FTP workflow overlooks the misalignment between the filtering and prediction objectives, failing to account for the additional errors that arise when using the approximated dynamics for prediction. The significant performance loss, stemming from model mismatch and objective misalignment, underscores the urgent need for more advanced techniques to effectively track time-varying channels with limited expert knowledge.

Data-driven methods leverage expressive neural networks, equipped with numerous trainable parameters, to learn complex channel dynamics directly from data without relying on expert knowledge \cite{kim2020massive, zhu2019adaptive, wei2022channel, jiang2022accurate}. These neural networks are trained to minimize prediction errors in an end-to-end manner, supervised by labeled CSI data. Empirical results demonstrate their superior ability to track fast-changing time series data using advanced network architectures, such as multilayer perceptrons (MLP) \cite{kim2020massive}, recurrent neural networks (RNN) \cite{zhu2019adaptive}, long short-term memory networks (LSTM) \cite{wei2022channel}, and Transformers \cite{jiang2022accurate}. However, without a principled integration of physical knowledge from the SSM and KF, data-driven methods often depend on extensive labeled CSI data for effective training, which is rarely accessible in practical scenarios. Furthermore, neural networks are typically treated as black boxes with limited interpretability \cite{shlezinger2023model}, raising concerns when deployed in security-critical V2X applications. Addressing these concerns will facilitate broader deployment of data-driven channel prediction methods for 5G/6G wireless networks in complex environments.

The inherent limitations of both model-based and data-driven channel prediction methods motivate their integration to leverage the benefits of both domains, posing the following fundamental questions:

\begin{itemize}
    \item[\textbf{Q1:}] Can the expressiveness of data-driven methods be integrated in a principled manner to enhance the prediction accuracy of model-based methods?
    \item[\textbf{Q2:}]  Conversely, can the label-free feature of model-based methods benefit data-driven methods?
    \item[\textbf{Q3:}] Is there any theoretical explanation underlying these mutual benefits?
\end{itemize}

\renewcommand{\arraystretch}{1.6}
\begin{table}[t]
    \centering
    \caption{Notation}
    \begin{tabular}{lp{0.7\columnwidth}}
        \hline
        \multicolumn{1}{c}{\textbf{Symbol}} & \multicolumn{1}{c}{\textbf{Meaning}} \\
        \hline
        \hline
        $a$ & Scalar, unbolded lowercase letter. \\
        $\mathbf{a}$ & Vector, bolded lowercase letter.\\
        $\mathbf{a}^\Re$, $\mathbf{a}^\Im$ & Real and imaginary parts of vector. \\
        $\mathbf{A}$ & Matrix, bolded uppercase letter.  \\
        $A_{n,m}$ & Matrix element in the $n$-th row and $m$-th column.\\
        $\mathbf{A}^\top$ & Transpose of matrix. \\
        $\mathbf{A}^H$ & Conjugate transpose of matrix.\\
        $\mathbf{A}^{\dagger}$ & Pseudo-inverse of matrix. \\
        $\det(\A)$ & Determinant of matrix. \\
        $\vec(\cdot)$ & Vectorization of matrix by column concatenation. \\
        $\nabla_{\mathbf{b}} f(\mathbf{b})$ &  Gradient of function $f(\mathbf{b})$ with respect to $\mathbf{b}$. \\
        $\|\cdot\|$ & $\ell_2$ norm. \\
        $\I_{n}$ &  $n \times n$ identity matrix. \\
        $\mathbf{0}_n$ &  $n \times n$ zero matrix. \\
        $\mathbf{0}_{m \times n}$ &  $m \times n$ zero matrix. \\
        $\mathbb{E}[\cdot]$ & Expectation operator. \\
        $\log_n(\cdot)$ & Logarithm to base $n$. \\
        $\{n\}$ & Index set containing ${1, 2, \ldots, n}$. \\
        $\bbN$ & Set of natural numbers. \\
        $\bbN^+$ & Set of positive integers. \\
        $\bbR^{+}$ & Set of positive real values. \\
        $\bbR^{m \times n}$ & Set of real values of dimensions $m \times n$. \\
        $\bbC^{m \times n}$ & Set of complex values of dimensions $m \times n$. \\
        $\cdot$ & Inner product. \\
        $\otimes$ & Kronecker product. \\
        $\mathcal{CN}(\boldsymbol{\mu}, \bSigma)$ & Complex multivariate Gaussian distribution with mean vector $\boldsymbol{\mu}$, and covariance matrix $\bSigma$. \\
        \hline
    \end{tabular}
    \label{table-notation}
\end{table}

\subsection{Contributions}
\label{sec-intro-contributions}
\noindent To answer these questions, this paper identifies the deviated Kalman gain as the root cause of the degraded prediction performance in the classical model-based FTP workflow. Consequently, we propose a hybrid model-based and data-driven FTP workflow, termed Kalman Prediction Integrated with Neural Network (KPIN), for accurately forecasting time-varying channels. Specifically, KPIN integrates a data-driven weighting matrix, learned by a neural network from data, into the FTP workflow to replace the deviated Kalman gain. Following this hybrid FTP workflow, the embedded neural network can be trained by maximizing the likelihood of historically received noisy signals, resulting in a novel unsupervised learning strategy that does not require labeled CSI data. For performance evaluation, various synthetic datasets are simulated based on a realistic channel model, \textit{3GPP\_TR\_38.901} \cite{3gpp_TS_38.901}, adhering to the latest standards set by the 3rd Generation Partnership Project (3GPP) and constraints summarized from practical scenarios. 

Our main contributions are as follows:

\begin{itemize}
    \item[\textbf{A1:}] We propose a novel high-accuracy prediction method for time-varying channels under high user mobility without expert knowledge, which utilizes a hybrid FTP workflow to effectively overcome model mismatch and objective misalignment.
    \item[\textbf{A2:}] An unsupervised learning strategy is developed to train KPIN without the need for labeled data, thereby circumventing the costly manual labeling process and lowering the barrier for practical deployment.
    \item[\textbf{A3:}] Theoretical analyses and ablation studies are provided to interpret the mutual benefits between the model-based and data-driven components of KPIN.
    \item[\textbf{A4:}] We evaluate KPIN across various scenario configurations, where it outperforms state-of-the-art methods that are either purely model-based or data-driven. KPIN demonstrates significant performance gain even under challenging practical conditions, such as substantial channel variations and high levels of signal noise. 
\end{itemize}

The remainder of the paper is organized as follows. Section \ref{sec-system} introduces the system model and problem formulation. Section \ref{sec-method} elaborates on the proposed KPIN method and theoretical analyses for answering the three fundamental questions listed before. Section \ref{sec-result} includes the experimental results. Section \ref{sec-conclusion} finally concludes this paper. The notation adopted in this paper is outlined in Table~\ref{table-notation}.

\section{Time-varying Channel Prediction}
\label{sec-system}
In this section, we first introduce the necessary background knowledge on the mmWave massive MIMO system in Section \ref{sec-system-background}, followed by our problem formulation for time-varying channel prediction in Section \ref{sec-system-problem}.

\subsection{Background}
\label{sec-system-background}

\subsubsection{Channel Model}
\begin{figure}[ht]
    \centering
    \includegraphics[width=\columnwidth]{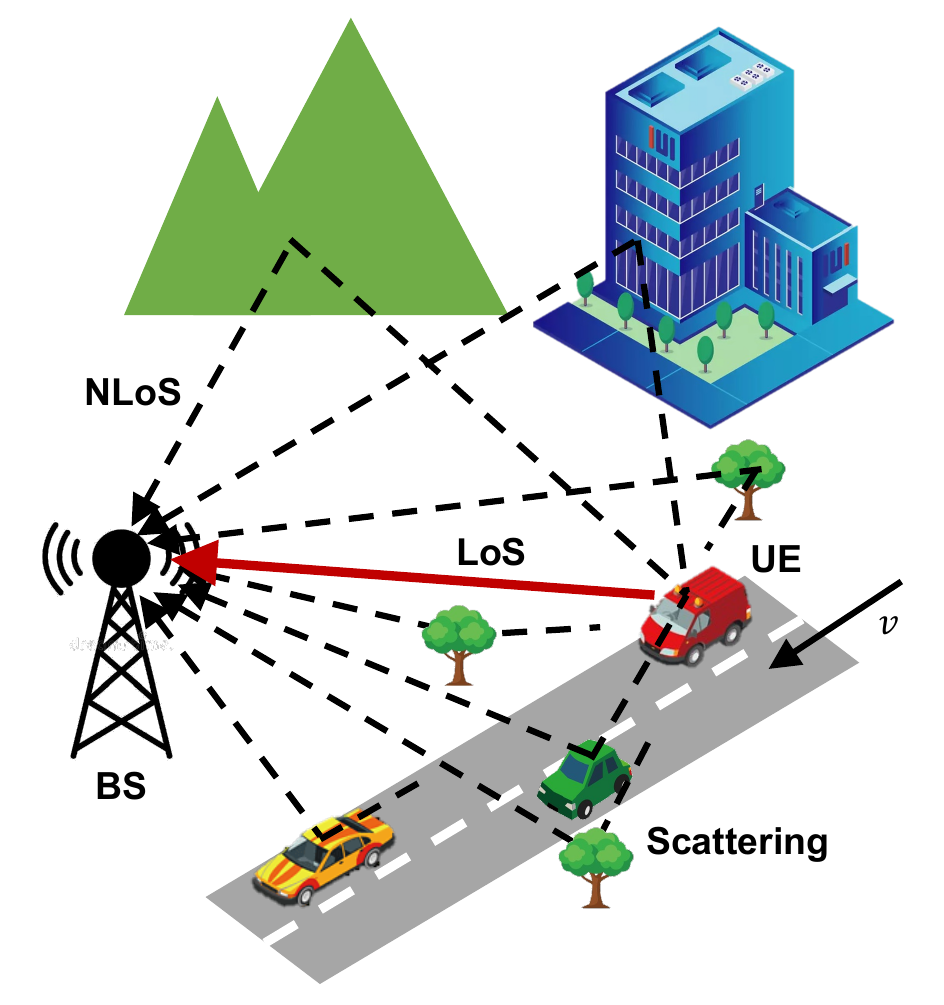}
    \caption{The schematic diagram of V2X scenario \cite{liao2019ekf}.}
    \label{fig-time-varying}
\end{figure}

\noindent As shown in Fig. \ref{fig-time-varying}, we consider uplink time-varying CSI prediction in a massive MIMO system operating in time-division duplex (TDD) mode \cite{goldsmith2005wireless}. The BS serves a single UE in an urban macro-cell environment with various obstacles, such as mountains, trees, buildings, and moving vehicles. The UE, equipped with a uniform planar array (UPA) of $M$ transmitter (Tx) elements, moves at a speed of $v$ km/h, while the BS is equipped with a UPA of $N$ receiver (Rx) elements. 

\begin{remark}
The assumed scenario is realistic and well covers various potential applications. Firstly, since CSI prediction for different UE is identical and independent, extending it to scenarios with multiple UE is straightforward \cite{kim2020massive}. Moreover, the uplink CSI prediction can aid in predicting the corresponding downlink CSI by leveraging the reciprocity of the TDD channel \cite{tse2005fundamentals}. Even for downlink CSI prediction in frequency-division duplex (FDD) mode, where channel reciprocity does not hold, the predicted uplink CSI can still be useful when incorporating partial reciprocity of the FDD channel \cite{tmc2018investigation}.   
\end{remark}

\begin{remark}
  The movement of the UE results in changes to its relative position with respect to the BS over time $t$, which in turn alters the propagation paths between the UE and the BS. As a result, the channels experienced by the moving UE change over time, a phenomenon referred to as \tb{time-varying channels} \cite{kim2020massive,jiang2022accurate}. 
\end{remark}

In the channel prediction task, at each time slot, $t \in \bbN$, the BS receives pilot signals from the UE and predicts the uplink CSI matrix $\H_{t+1} \in \bbC^{N \times M}$ for the subsequent time slot $(t+1)$:
\begin{equation}
    \H_{t+1}=\left[\begin{array}{ccc}
        H_{t+1,1,1} & \cdots &  H_{t+1,1,M} \\
        \vdots & \ddots & \vdots \\
        H_{t+1,N,1} & \cdots &  H_{t+1,N,M}
    \end{array}\right],
\end{equation}
where each entry $H_{t+1,n,m} \in \bbC$ denotes the CSI for each Rx-Tx pair $\{n,m\}$, $1 \leq n \leq N$, $1 \leq m \leq M$.  $H_{t+1,n,m}$ is further defined as the weighted sum of the line-of-sight (LoS) component $H_{t,n,m}^{\LoS} \in \mathbb{C}$ and the non-line-of-sight (NLoS) component $H_{t,n,m}^{\NLoS} \in \mathbb{C}$, which accounts for paths such as reflections, diffractions, and scattering. Additional modeling factors include oxygen absorption, blockage effects, etc.

In this work, we assume that the wireless channels follow the latest channel model, \textit{3GPP\_TR\_38.901} \cite{3gpp_TS_38.901}, which offers several key advantages ideal for simulating the targeted dynamic scenarios. The channel model we consider can be adapted to earlier channel models, such as 3D SCM \cite{3gpp_TS_36.873} and IMT-Advanced \cite{ITU-RM.2135}, and supports a wide carrier frequency range from $f=0.5$ to $100$ GHz, with special considerations for mmWave characteristics. Furthermore, the selected channel model supports large antenna arrays and accommodates user mobility, with a speed up to $v=500$ km/h.  For further details on this channel model, the interested reader is referred to \cite{3gpp_TS_38.901}, as these specifics are omitted here due to page limitations.

\subsubsection{Dynamic Condition}
\label{sec-system-dynamics}
For notational brevity, the duration of each time slot has been normalized to unity with respect to the pilot signal period $\Delta t \in \bbR^+$. This normalization ensures the time index $t$ an integer ($t \in \bbN$). The choice of $\Delta t$ [ms] is determined by the following criterion:
\begin{equation}
    \Delta t \triangleq k \cdot T_c,
    \label{eq-ssm-delta}
\end{equation}
where $k \in \mathbb{R}^+$ is an adjustable dynamic condition that accounts for channel variations, and $T_c$ [ms] represents the channel coherence time, during which the channels can be reasonably considered time-invariant \cite{marzetta2016fundamentals}:
\begin{equation}
    T_c  = \frac{540}{v  \cdot f }   ,
    \label{eq-coherence}
\end{equation}
which is inversely propotional to the UE speed $v$ [km/h] and the carrier frequency $f$ [GHz]. 

\begin{remark} 
  \label{remark-k}
  A larger $k$ indicates a longer time slot $\Delta t$, resulting in decreased correlation between channel states at consecutive time steps and implying more severe channel aging. Since next-generation wireless communication relies on increasingly higher frequencies, it is essential to consider the Doppler effects resulting from both the UE velocity $v$ and carrier frequency $f$, rather than focusing solely on $v$ as in previous works \cite{kim2020massive,jiang2022accurate}. Hence, the dynamical condition $k$ provides a more comprehensive metric to measure how quickly the channels become outdated. 
\end{remark}

\subsubsection{Signal Model}
\label{sec-system-transmission}
\noindent In TDD mode, the UE sends the length-$\tau$ ($\tau \in \bbN^+$) pilot matrix $\Q_{\text{pilot}} \in \bbC^{M \times \tau}$ to the BS for the uplink CSI estimation. The following semi-unitary pilot matrix with the constraint $M \leq \tau$ is adopted:
\begin{equation}
    \Q_{\text{pilot}} \cdot \Q_{\text{pilot}}^H =\tau \cdot \I_{M},
    \label{eq-semi}
\end{equation}
which forces the pilot signals assigned to different Tx elements to be orthogonal to each other, making the CSI estimation for each Tx element independent \cite{tse2005fundamentals}. 

The received signals $\Y_t \in \bbC^{N \times \tau}$ at the BS can then be written in the vectorized form as follows: 
\begin{equation}
    \y_t=\Q \cdot \h_t+\v_t, 
    \label{eq-transmission-vec}
\end{equation}
where  $\y_t=\operatorname{vec}(\Y_t) \in \mathbb{C}^{ \tau N} $ and $\h_t=\operatorname{vec}(\H_t) \in \bbC^{MN}$ are vectorized versions of the received signals and CSI matrix, respectively. The transformed pilot matrix for dimension alignment with the vectorized CSI matrix is given by:
\begin{equation}
    \Q=\left(\sqrt{\rho} \Q_{\text{pilot}}^{\top} \otimes \I_{N} \right)
    \in \bbC^{\tau N \times MN},
    \label{eq-pilot-transformed}
\end{equation}
where $\rho \in \mathbb{R}^+$ controls the Tx power for the desired signal-to-noise ratio (SNR). $\v_t  \sim \mathcal{CN}(\b0_{\tau N \times 1}, \bSigma_{\v}) $ represents an independent and identically distributed (i.i.d.) zero-mean additive white Gaussian noise (AWGN) with covariance matrix $\bSigma_{\v}=\sigma_{\v}^2 \cdot \I_{\tau N}
$, where $\sigma_{\v} \in \bbR^+$ is the standard deviation. Note that the signal noise $\mathbf{v}_{t^{\prime}}$ ($t^{\prime} \neq t$) is not related, thus uncorrelated, to the channel $\h_t$:
\begin{equation}
    \mathbb{E}\left[\mathbf{h}_t \cdot \mathbf{v}_{t^{\prime}}^H\right] = \mathbb{E}\left[\mathbf{h}_t\right] \cdot \mathbb{E}\left[\mathbf{v}_{t^{\prime}}\right]^H
    =  \mathbf{0}_{MN \times \tau N}.
    \label{eq-signal_noise}
\end{equation}

\subsection{Problem Formulation}
\label{sec-system-problem}
\noindent In the channel prediction task, the BS predicts the uplink CSI vector $\h_{t+1}$ based on previously received noisy signals $\y_{t-p+1:t} \triangleq [\y_{t-p+1}^\top, \ldots,\y_{t}^\top]^\top $, where $p \in \bbN^+$ denotes the complexity order of the task \cite{kim2020massive}. Formally, the optimization problem can be formulated as:

\begin{subequations}
    \begin{align}
    & \min_{s} \ \bbE \left[ \norm {\h_{t+1}-\hh_{t+1 \mid t}}^2\right],\\
    & \operatorname{s.t.}\quad \hh_{t+1 \mid t}=s \left(\y_{t-p+1:t}\right),
    \end{align}
    \label{eq-objective-predict}
\end{subequations}

\noindent where $\hh_{t+1 \mid t}$ is the predicted CSI vector given the received signals up to the time step $t$. The mapping function $s (\cdot)$ denotes an arbitrary prediction method, such as a model-based FTP workflow or a data-driven Transformer. Our objective is to integrate model-based and data-driven channel prediction methods in a principled manner to leverage their mutual benefits, as detailed in Section \ref{sec-method}.

\begin{remark}
    \label{remark_only}
  In practical scenarios, we only observe a single sequence of historically received noisy signals with length $T \in \bbN^+$, denoted by $\{\mathbf{y}_t\}_{t=1}^T$. It is important to note that the input for channel prediction (\ref{eq-objective-predict}) consists solely of the noisy signals $\y_{t-p+1:t}$. In contrast, some previous methods utilize the past perfect channels  $\h_{t-p+1:t} \triangleq [\h_{t-p+1}^\top, \ldots,\h_{t}^\top]^\top $ as input (see (12) in \cite{jiang2022accurate}), which is not feasible in practical scenarios.
\end{remark}

Based on the predicted CSI vector $\hh_{t+1 \mid t}$, the achievable rate $r \in \bbR^+ $ for downlink data transmission is computed under the assumption that Gaussian symbols are transmitted \cite{marzetta2016fundamentals}:
\begin{equation}
    r =\bbE \left[\log _2  \det(\I_M+\frac{\rho}{M \cdot \sigma_\v^2} \P \cdot \hH_{t+1 \mid t} \cdot \hH_{t+1 \mid t}^H \cdot \P^H) \right],
    \label{eq-rate}
\end{equation}
where $\hh_{t+1 \mid t} = \vec(\hH_{t+1 \mid t})$. The zero-forcing (ZF) precoder matrix $\P \in \bbC^{M \times N} $ adopted at the BS is calculated as:
\begin{equation}
    \P=\left(\hH_{t+1 \mid t}^H \cdot \hH_{t+1 \mid t}\right)^{-1} \cdot \hH_{t+1 \mid t}^H.
    \label{eq-zf}
\end{equation}

\section{Kalman Prediction Integrated with Neural Network}
\label{sec-method}
\noindent This section introduces KPIN, which replaces the deviated Kalman gain in the classical model-based FTP workflow with a data-driven weighting matrix, trained according to the maximum likelihood criterion without requiring labeled data. Section \ref{sec-method-ssm} outlines how to construct the model-based SSM. Section \ref{sec-method-workflow} analyzes the problems with the conventional model-based FTP workflow. Section \ref{sec-method-kpin} to \ref{sec-method-discuss} address the three fundamental questions raised in Section \ref{sec-intro}, respectively.

\subsection{Model-based SSM}
\label{sec-method-ssm}
\noindent The well-established SSM offers a principled framework for general dynamic processes, comprising a transition model and an observation model \cite{sarkka2023bayesian}. The transition model describes how the latent state evolves with time. The observation model maps the latent state to the observation. In the context of channel prediction, the CSI vector $\h_t$ acts as the latent state, and the received noisy signal vector $\y_t$ serves as the observation.

To construct the transition model, the AR model \cite{haykin2002adaptive} is utilized to approximate the channel dynamics, as it allows closed-form parameter estimation \cite{kim2020massive} and outperforms linear extrapolation \cite{yin2020addressing} in terms of expressiveness. Unlike methods based on the sum-of-sinusoids \cite{wong2005joint}, the AR model maintains a linear structure with AWGN, facilitating the proposed hybrid FTP workflow through a linear Gaussian SSM. In addition, the AR model delivers comparable performance to the ARMA model in the FTP workflow while being computationally more efficient \cite{kashyap2017performance}. It characterizes time-varying channels as different realizations of the same dynamic process \cite{haykin2002adaptive}:
\begin{equation}
    \h_t = \sum_{j=1}^p \bPhi_j \cdot \h_{t-j}+\u_t,
    \label{eq-ar}
\end{equation}
where the AR order $p$ is consistent with the complexity order in (\ref{eq-objective-predict}). $\bPhi=\left[\bPhi_1, \bPhi_2, \hdots, \bPhi_p\right]\in \bbC^{MN \times pMN}$ comprises all AR coefficient matrices, $\bPhi_j \in \bbC^{MN \times MN}, \forall j \in \{p\}$. The noise $\u_{t} \sim \mathcal{CN}(\b0_{MN \times 1}, \bSigma_{\u}) $ is assumed to be i.i.d. AWGN. Hence, the future noise $\mathbf{u}_{t^{\prime}}$ ($t^{\prime}>t$) is not related, thus uncorrelated, to channels $\left\{\mathbf{h}_t\right\}_{t=1}^{t^{\prime}-1}$ at previous time steps:
\begin{equation}
    \mathbb{E}\left[\mathbf{h}_t \cdot \mathbf{u}_{t^{\prime}}^H\right] = \  \mathbb{E}\left[\mathbf{h}_t\right] \cdot \mathbb{E}\left[\mathbf{u}_{t^{\prime}}\right]^H
    =  \mathbf{0}_{MN}.
    \label{eq-variable-noise}
\end{equation}

The linear structure and i.i.d. AWGN condition in (\ref{eq-transmission-vec}) and (\ref{eq-ar}) provide an effective framework for analytically estimating the AR parameters $\bPhi$ and $\bSigma_{\u}$ from historically received noisy signals $\{ \y_t\}_{t=1}^{T}$, based on Propositions \ref{pro-yw} and \ref{pro-yw-empirical}:
\begin{subequations}
    \begin{align}
        \boldsymbol{\Phi}^H & \approx ({\hC_{\text{all}} + \epsilon \cdot \I_{pMN}})^{-1} \cdot \hC,
        \label{eq-yw-empirical}\\   
        \boldsymbol{\Sigma}_{\u} & \approx \hC_0 - \hC^H \cdot \boldsymbol{\Phi}^H,
        \label{eq-yw-empirical-ut}
    \end{align}
\end{subequations}

\noindent where $\hC_{\text{all}}$, $\hC$ and $\hC_0$ are approximations of $\C_{\text{all}}$, $\C$ and $\C_0$ as defined in Proposition \ref{pro-yw}, relying on the definition of $\hC_k$ in Proposition \ref{pro-yw-empirical}. $\epsilon \in \bbR^+$ is a small perturbation factor added to avoid numerical issues, as $\hC_{\text{all}} \in \bbC^{pMN \times pMN}$ can become ill-conditioned when the AR order $p$ is high \cite{baddour2005autoregressive}.

\begin{proposition}
    Assuming a linear structure and i.i.d. AWGN in the approximated channel dynamics (\ref{eq-ar}), we have the following Yule-Walker equations \cite{yule1971method,walker1931periodicity}:
    \begin{subequations}
        \begin{align}
            \C_{\text{all}} \cdot \boldsymbol{\Phi}^H & = \C, \label{yw_1} \\
            \C^H \cdot \boldsymbol{\Phi}^H + \boldsymbol{\Sigma}_{\u} & = \C_0, \label{yw_2}
        \end{align}
    \end{subequations}

    \noindent where the right-hand side (RHS) matrix $\C$ and the coefficient matrix $\C_{\text{all}}$ are defined in (\ref{eq-C-vec_main}) and (\ref{eq-C-all-even_main}), respectively, based on the auto-covariance matrix $\mathbf{C}_k$ defined in (\ref{eq-auto-lag_main}).

    \label{pro-yw}
\end{proposition}

\begin{italicproof}
    
\noindent \tb{Proof.} We first present the notion of the auto-covariance matrix and its properties, which facilitate the subsequent derivation of (\ref{yw_1}) and (\ref{yw_2}), respectively, based on the different cases of $k$.

It is usually assumed that the auto-covariance matrix $\mathbf{C}_k  \in \mathbb{C}^{MN \times MN}$ between two arbitrary channel vectors is invariant to a time shift $\Delta t \in \mathbb{Z}$ \cite{haykin2002adaptive}:
    \begin{equation}
            \mathbf{C}_k \triangleq  \mathbb{E}\left[\mathbf{h}_{t-k} \cdot \mathbf{h}_{t}^H\right] 
            = \mathbb{E}\left[\mathbf{h}_{t+\Delta t-k} \cdot {\mathbf{h}_{t+\Delta t}^H}\right],
            \label{eq-auto-lag_main}
    \end{equation}
    which implies that $\mathbf{C}_k $ depends on only their relative time lag $k \in \mathbb{Z}$, instead of the absolute time step $(t-k)$ or $t$. In addition, it is straightforward to derive the evenness of the auto-covariance matrix:
    \begin{subequations}
        \begin{align}
            \mathbf{C}_{-k} \triangleq \ & \mathbb{E}\left[\mathbf{h}_{t-(-k)} \cdot \mathbf{h}_{t}^H\right]\\
            = \ & \mathbb{E}\left[\mathbf{h}_t \cdot \mathbf{h}_{t+k}^H\right]^H \\
             \overset{\text{(\ref{eq-auto-lag_main})}}{=} \ &\mathbb{E}\left[\mathbf{h}_{t-k} \cdot \mathbf{h}_t^H\right]^H \triangleq \mathbf{C}_k^H.
        \end{align}
        \label{eq_eveness}
    \end{subequations}

    \begin{enumerate}
        \item The case $ 0<k\leq p$ for proving (\ref{yw_1}):
        
        Making conjugate transpose of ($\ref{eq-ar}) $ and left-multiplying it with $ \mathbf{h}_{t-k}$ on both sides yield:
        \begin{equation}
            \mathbf{h}_{t-k} \cdot \mathbf{h}_t^H=\sum_{j=1}^p \mathbf{h}_{t-k} \cdot \mathbf{h}_{t-j}^H \cdot \boldsymbol{\Phi}_j^H+\mathbf{h}_{t-k} \cdot \mathbf{u}_t^H.
            \label{eq-left-multiply_main}
        \end{equation}
        
        Taking expectation on both sides of ($\ref{eq-left-multiply_main}$) gives:
    \begin{subequations}
        \begin{align}
            \quad & \mathbb{E}\left[\mathbf{h}_{t-k} \cdot \mathbf{h}_t^H\right] \label{eq-expect-start_main}\\
            \overset{(\ref{eq-left-multiply_main})}{=} & \mathbb{E} \left[\sum_{j=1}^p \mathbf{h}_{t-k} \cdot \mathbf{h}_{t-j}^H \cdot \boldsymbol{\Phi}_j^H+\mathbf{h}_{t-k} \cdot \mathbf{u}_t^H\right] \\
             \overset{(\ref{eq-variable-noise})}{=} & \sum_{j=1}^p \mathbb{E}\left[\mathbf{h}_{t-k} \cdot \mathbf{h}_{t-j}^H \right]\cdot \boldsymbol{\Phi}_j^H\\
            \overset{\text{(\ref{eq-auto-lag_main})}}{=}  & \sum_{j=1}^p \mathbb{E}\left[\mathbf{h}_{t-(k-j)} \cdot \mathbf{h}_{t}^H\right] \cdot \boldsymbol{\Phi}_j^H ,\label{eq-expect-end_main}
        \end{align}
        \label{eq-expect}
    \end{subequations}

    \noindent where the equivalence between (\ref{eq-expect-start_main}) and (\ref{eq-expect-end_main}) can be rewritten compactly as:
    \begin{equation}
        \mathbf{C}_k=\sum_{j=1}^p \mathbf{C}_{k-j} \cdot \boldsymbol{\Phi}_j^H .
        \label{eq-ck-equivalence_main}
    \end{equation}

    Then, by ergodically unrolling (\ref{eq-ck-equivalence_main}) for $0<k\leq p $, the resulting $p$ equations can be gathered to form the following linear system \cite{yule1971method,walker1931periodicity}:
    \begin{equation}
        \mathbf{C}_{\text{all}} \cdot \boldsymbol{\Phi}^{H} =\mathbf{C},
        \label{eq-yw-proof}
    \end{equation}
    where $\C$ and $\C_{\text{all}}$ are defined as:
    \begin{equation}
        \C \triangleq [
        \C_1^\top, \C_2^\top, \dots,  \C_p^\top]^\top,
        \label{eq-C-vec_main}
    \end{equation}

    \vspace*{-1em}

     \begin{equation}
        \setlength{\arraycolsep}{0.04pt} 
        \renewcommand{\arraystretch}{1.2} 
        \mathbf{C}_{\text{all}} \triangleq \left[\begin{array}{cccccc}
        \mathbf{C}_0 & \mathbf{C}_{-1} & \cdots & \mathbf{C}_{1-p} \\
        \mathbf{C}_1 & \mathbf{C}_0 & \cdots & \mathbf{C}_{2-p} \\
        \vdots & \vdots & \ddots & \vdots \\
        \mathbf{C}_{p-1} & \mathbf{C}_{p-2} & \cdots & \mathbf{C}_0
        \end{array}\right]
        \overset{(\ref{eq_eveness})}{=}
        \left[\begin{array}{cccccc}
        \mathbf{C}_0 & \mathbf{C}_1^H & \cdots & \mathbf{C}_{p-1}^H \\
        \mathbf{C}_1 & \mathbf{C}_0 & \cdots & \mathbf{C}_{p-2}^H \\
        \vdots & \vdots & \ddots & \vdots \\
        \mathbf{C}_{p-1} & \mathbf{C}_{p-2} & \cdots & \mathbf{C}_0
        \end{array}\right].
        \label{eq-C-all-even_main}
    \end{equation}
    
    \item The case $ k=0$ for proving (\ref{yw_2}):
    
    Making conjugate transpose of ($\ref{eq-ar}) $ and left-multiplying it with $ \mathbf{h}_{t}$ on both sides yield:
    \begin{equation}
        \mathbf{h}_{t} \cdot \mathbf{h}_t^H=\sum_{j=1}^p \mathbf{h}_{t} \cdot \mathbf{h}_{t-j}^H \cdot \boldsymbol{\Phi}_j^H+\mathbf{h}_{t} \cdot \mathbf{u}_t^H.
        \label{eq-left-multiply_0}
    \end{equation}

    Taking expectation on both sides of ($\ref{eq-left-multiply_0}$) gives:
    \allowdisplaybreaks
    \begin{subequations}
        \begin{align}
            &\mathbb{E}\left[\mathbf{h}_{t} \cdot \mathbf{h}_t^H\right] \label{eq-expect-zero-start_main} \\
            \overset{(\ref{eq-left-multiply_0})}{=} & \mathbb{E} \left[\sum_{j=1}^p \mathbf{h}_{t} \cdot \mathbf{h}_{t-j}^H \cdot \boldsymbol{\Phi}_j^H+\mathbf{h}_{t} \cdot \mathbf{u}_t^H\right] \\
            \overset{(\ref{eq-ar})}{=} & \sum_{j=1}^p  \mathbb{E}\left[\mathbf{h}_{t} \cdot \mathbf{h}_{t-j}^H \right] \cdot \boldsymbol{\Phi}_j^H +\sum_{j=1}^p  \boldsymbol{\Phi}_j \cdot \mathbb{E}\left[ \mathbf{h}_{t-j} \cdot \mathbf{u}_t^H\right] \label{eq-expect-zero-phi-2}  \notag \\
            &  + \mathbb{E}\left[ \mathbf{u}_t\cdot \mathbf{u}_t^H\right] \\
            \overset{(\ref{eq-variable-noise})}{=}  & \sum_{j=1}^p  \mathbb{E}\left[\mathbf{h}_{t} \cdot \mathbf{h}_{t-j}^H \right] \cdot \boldsymbol{\Phi}_j^H +\mathbb{E}\left[ \mathbf{u}_t\cdot \mathbf{u}_t^H\right]\\
            \overset{\text{(\ref{eq-auto-lag_main})}}{=}  & \sum_{j=1}^p  \mathbb{E}\left[\mathbf{h}_{t-(-j)} \cdot \mathbf{h}_{t}^H\right] \cdot \boldsymbol{\Phi}_j^H+\mathbb{E}\left[ \mathbf{u}_t\cdot \mathbf{u}_t^H\right],
             \label{eq-expect-zero-end_main}
        \end{align}
        \label{eq-expect-zero}
    \end{subequations}

    where the equivalence between (\ref{eq-expect-zero-start_main}) and (\ref{eq-expect-zero-end_main}) can be rewritten compactly as:
    \begin{subequations}
    \begin{align}
    \mathbf{C}_0= \ & \sum_{j=1}^p \mathbf{C}_{-j} \cdot \boldsymbol{\Phi}_j^H + \boldsymbol{\Sigma}_{\u}\\
    \overset{(\ref{eq_eveness})}{=} & \sum_{j=1}^p\mathbf{C}_{j}^H \cdot \boldsymbol{\Phi}_j^H \  + \boldsymbol{\Sigma}_{\u}\\
    \overset{(\ref{eq-C-vec_main})}{=}  &  \mathbf{C}^H \cdot \boldsymbol{\Phi}^H   + \boldsymbol{\Sigma}_{\u}.
    \end{align}
\end{subequations} \hfill $\blacksquare$

\end{enumerate}   
\end{italicproof}

\begin{proposition}
    \label{pro-yw-empirical}
    Assuming a linear structure and i.i.d. AWGN in the signal model (\ref{eq-transmission-vec}), $\mathbf{C}_k$ can be approximated by $\hat{\mathbf{C}}_k$ based on $\left\{ \y_t \right\}_{t=1}^{T}$ as:
    \begin{equation}
        \begin{aligned}
          \hC_k
        =   \left\{\begin{array}{lll}
            \Q^{\dagger} \cdot(\frac{1}{T-1} \sum_{t=k+1}^{T} \y_{t-k} \cdot \y_t^H) \cdot (\Q^H)^{\dagger}, 0<k \leq p,\\
            \Q^{\dagger} \cdot(\frac{1}{T-1} \sum_{t=k+1}^{T} \y_{t} \cdot \y_t^H-\boldsymbol{\Sigma}_{\v}) \cdot (\Q^H)^{\dagger}, k=0.
        \end{array}\right. 
        \end{aligned}
        \label{ck_approx}
    \end{equation}    
\end{proposition}

\begin{italicproof}
    \noindent \tb{Proof.} We firstly derive the auto-covariance matrix of $\mathbf{y}_{t}$, defined as $\mathbb{E}\left[\mathbf{y}_{t-k} \cdot \mathbf{y}_t^H\right]$, with a time lag of \(k\) (where \(k \in \mathbb{Z}\) and \(0 \leq k \leq p\)). Following this, we introduce methods to approximate this term, which serves as the foundation for our goal: the approximation of \(\mathbf{C}_k\).

    \begin{subequations}
        \begin{align}
        & \mathbb{E}\left[\mathbf{y}_{t-k} \cdot \mathbf{y}_t{ }^H\right]\label{eq-auto-y-start_main} \\
        \overset{(\ref{eq-transmission-vec})}{=} \  & \Q \cdot \mathbb{E}\left[\mathbf{h}_{t-k} \cdot \mathbf{h}_t^H\right] \cdot \Q^H+\Q \cdot \bbE\left[ \mathbf{h}_{t-k} \cdot \mathbf{v}_t^H\right] \notag \\
        & +\bbE\left[\mathbf{h}_t \cdot \mathbf{v}_{t-k}^H  \right]^H\cdot \Q^H+\bbE\left[\mathbf{v}_{t-k} \cdot \mathbf{v}_t{ }^H\right] \label{eq-expect-Q}\\
        \overset{(\ref{eq-signal_noise})}{=}  \ &\Q \cdot \mathbb{E}\left[\mathbf{h}_{t-k} \cdot \mathbf{h}_t^H\right] \cdot \Q^H+\bbE\left[\mathbf{v}_{t-k} \cdot \mathbf{v}_t{ }^H\right] \\
        = \ & \left\{\begin{array}{lll}
            &\Q \cdot \mathbb{E}\left[\mathbf{h}_{t-k} \cdot \mathbf{h}_t^H\right] \cdot \Q^H&, 0<k \leq p \\
            &\Q \cdot \mathbb{E}\left[\mathbf{h}_{t} \cdot \mathbf{h}_t^H\right] \cdot \Q^H+\boldsymbol{\Sigma}_{\v} &, k=0\\
        \end{array}\right.\label{eq-ck_iid} \\
        \overset{(\ref{eq-auto-lag_main})}{=} & \left\{\begin{array}{lll}
            &\Q \cdot \mathbf{C}_k  \cdot \Q^H&, 0<k \leq p, \\
            &\Q \cdot \mathbf{C}_0 \cdot \Q^H+\boldsymbol{\Sigma}_{\v} &, k=0.\\
        \end{array}\right.
        \label{eq-auto-y-end_main}
        \end{align}
        \label{eq-auto-y}
    \end{subequations}

    \noindent where the transition to (\ref{eq-ck_iid}) holds due to the i.i.d. assumption in the signal model (\ref{eq-transmission-vec}).

    Then, from the equivalence of (\ref{eq-auto-y-start_main}) and (\ref{eq-auto-y-end_main}), we can recover $\mathbf{C}_k$ from the received noisy signals as:
\begin{equation}
     \mathbf{C}_k =   \left\{\begin{array}{lll}
    \Q^{\dagger} \cdot \mathbb{E}\left[\mathbf{y}_{t-k} \cdot \mathbf{y}_t^H\right] \cdot\left(\Q^H\right)^{\dagger} &, 0<k \leq p, \\
    \Q^{\dagger}  \cdot \left(\mathbb{E}\left[\mathbf{y}_{t} \cdot \mathbf{y}_t^H\right]-\boldsymbol{\Sigma}_{\v}\right) \cdot\left(\Q^H\right)^{\dagger} & , k=0.
    \end{array}\right.
    \label{eq-auto-h-recover}
\end{equation}

However, as illustrated in Remark \ref{remark_only}, there is no means to accurately compute $\mathbb{E}\left[\mathbf{y}_{t-k} \cdot \mathbf{y}_t^H\right]$, since we only have the received noisy signals with a finite length of $T$. Therefore, the auto-covariance matrix of $\y_t$ can be empirically estimated over $\left\{\y_t\right\}_{t=1}^T$:
\begin{equation}
    \begin{aligned}
    \bbE \left[\y_{t-k} \cdot \y_t^H\right] 
    \approx \ &  \left\{\begin{array}{lll}
        \frac{1}{T-1} \sum_{t=k+1}^{T} \y_{t-k} \cdot \y_t^H &, 0<k \leq p, \\
        \frac{1}{T-1} \sum_{t=1}^{T}\y_{t} \cdot \y_t^H &, k=0.\\
    \end{array}\right.  
    \end{aligned}
    \label{eq-auto-y-approx}
\end{equation}
\noindent which implies that $\mathbf{C}_k$ can be approximated by the empirical auto-covariance matrix $\hat{\mathbf{C}}_k$ in (\ref{ck_approx}), based on the equivalence in (\ref{eq-auto-h-recover}) and the approximation of $\mathbb{E}\left[\mathbf{y}_{t-k} \cdot \mathbf{y}_t^H\right]$ in (\ref{eq-auto-y-approx}). 

\hfill $\blacksquare$
\end{italicproof}

\begin{figure}[t]
    \centering
    \includegraphics[width=\columnwidth]{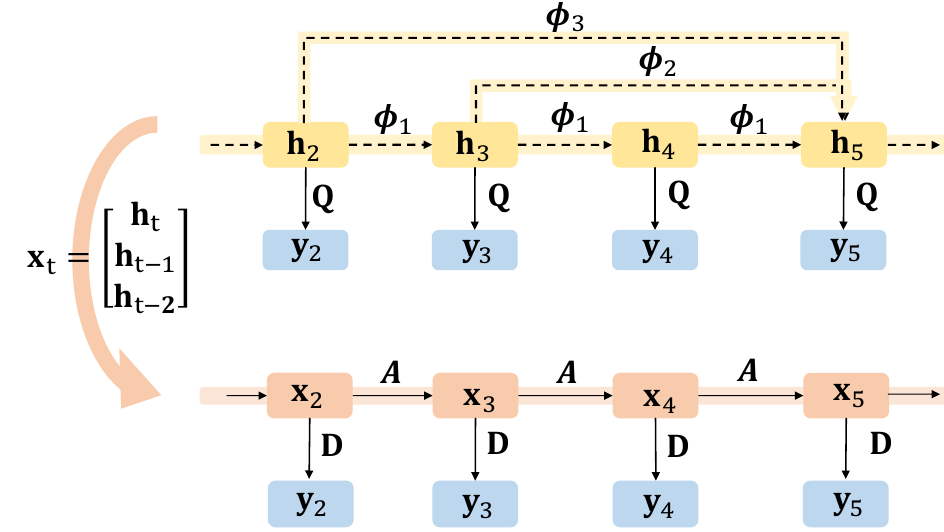}
    \caption{Latent state augmentation with $p=3$.}
    \label{fig-markov}
\end{figure}

Then, the AR model of order $p$ needs to be transformed into a first-order Markovian transition model \cite{sarkka2023bayesian}, facilitating the proposed FTP workflow. As illustrated in Fig. \ref{fig-markov} with $p=3$ being a toy example, a common method involves augmenting the latent state vector with previous CSI vectors. The augmented latent state is denoted as $\x_t \triangleq [\h_t^\top, \ldots,\h_{t-p+1}^\top]^\top \in \bbC^{pMN}$, and the transition model is then given by:
\begin{equation}
    \x_t=  \A \cdot \x_{t-1} +\B \cdot \u_{t},
    \label{eq-ssm-transition}
\end{equation}
where $\A=[\bPhi;\I_{(p-1)MN},\b0_{(p-1)MN \times MN}] \in \bbC^{pMN \times pMN}$ acts as the transition matrix, and the coefficient matrix $\B = [\I_{MN},\b0_{MN \times (p-1)MN}^\top]^\top \in \bbC^{pMN \times MN }$ is introduced for dimension alignment. 

In correspondence with the augmented latent state, the observation model is derived from the signal model (\ref{eq-transmission-vec}):
\begin{equation}
    \y_t= \D \cdot \x_t  + \v_t,
    \label{eq-ssm-observation}
\end{equation}
where the observation matrix $\D \in \bbC^{\tau N \times pMN}$ is given by:
\begin{equation}
    \D=\Q \cdot \B^{\top}.
    \label{eq-observation-matrix}
\end{equation}

\begin{algorithm}[t]
    \caption{SSM Construction}
     \begin{algorithmic} 
        \State \textbf{Inputs:} Historically received noisy signals $\{ \y_t\}_{t=1}^{T}$
    \end{algorithmic}
    \begin{algorithmic}[1] 
            \State Compute an approximation of $\mathbb{E}\left[\mathbf{y}_{t-k} \cdot \mathbf{y}_t{ }^H\right]$ as (\ref{eq-auto-y-approx})
            \State Compute an approximation of $\C_k$ as (\ref{ck_approx})
            \State Obtain an estimate of $\bPhi$ as (\ref{eq-yw-empirical})
            \State Obtain an estimate of $\bSigma_{\u}$ as (\ref{eq-yw-empirical-ut})
            \State Construct the transition function as (\ref{eq-ssm-transition})
            \State Construct the observation function as (\ref{eq-ssm-observation})
    \end{algorithmic}
    \begin{algorithmic} 
        \State \textbf{Outputs:} SSM parameters as (\ref{eq-ssm-aug})
    \end{algorithmic}
    \label{alg-ssm}
\end{algorithm}

Eventually, the transition and observation models in (\ref{eq-ssm-transition}) and (\ref{eq-ssm-observation}) are nested to be a linear Gaussian SSM, as visualized at the bottom of Fig. \ref{fig-markov} and as outlined in Algorithm \ref{alg-ssm}:
\begin{equation}
    \left\{
        \begin{array}{llll}
        \x_{t}= \ \ \A \cdot \x_{t-1} & + \ \ \mathbf{B} \cdot \u_{t} & \quad \text{(transition)},\\
        \y_t= \quad \D \cdot \x_t  & + \quad \v_t & \quad \text{(observation)}. 
        \end{array} 
    \right.
    \label{eq-ssm-aug}
\end{equation}

\subsection{Problems with Model-based FTP Workflow}
\label{sec-method-workflow}

\noindent This subsection identifies critical issues in the conventional model-based FTP workflow, motivating the development of the proposed hybrid FTP workflow in Section \ref{sec-method-kpin}.

The conventional model-based FTP workflow involves a recursive process of filtering and prediction over time. At each time step $t$, the prior estimate $\hat{\mathbf{x}}_{t \mid t-1}$ represents the predicted value of $\mathbf{x}_t$ given the received signals up to the time step $(t-1)$, while the posterior estimate $\hat{\mathbf{x}}_{t \mid t}$ denotes the filtered value of $\mathbf{x}_t$ given the received signals up to the time step $t$. The details are as follows:
\begin{enumerate}
    \item \textbf{Filter:}
    
    Based on the prior estimate $\hx_{t \mid t-1}$, its covariance matrix $ \P_{t \mid t-1} \in \bbC^{pMN \times pMN}$, the predicted signal $\hy_{t \mid t-1}$ and its covariance matrix $\S_{t \mid t-1}\in \bbC^{\tau N \times \tau N}$, KF is implemented to yield the posterior estimate $\hx_{t \mid t}$ and its covariance matrix $\P_{t \mid t} \in \bbC^{pMN \times pMN}$:
    \begin{subequations}
        \begin{align}
            \hx_{t \mid t}& = \hx_{t \mid t-1}+ \K_t\cdot \Delta \y_t, \label{eq-predictor-filtering-kt} \\  
            \P_{t \mid t} & = \P_{t \mid t-1}  - \K_t \cdot \S_{t \mid t-1}\cdot \K_t^H,
        \end{align}	
        \label{eq-predictor-filtering}
    \end{subequations}

    \noindent where $\K_t \in \bbC^{pMN \times \tau N}$ is the Kalman gain and $\Delta \y_t$ denotes the difference between the newly received signal $\y_t$ and the predicted signal $\hy_{t \mid t-1}$:

    \begin{subequations}
        \begin{align}
            \K_t & =\P_{t \mid t-1} \cdot \D^H \cdot \S_{t \mid t-1}^{-1} \label{eq-kg}, \\  
            \Delta \y_t & =\y_t-\hy_{t \mid t-1}. 
        \end{align}	
    \end{subequations}

    \item \textbf{Predict:}
    
    Next, the posterior estimate $\hx_{t \mid t}$ passes through the transition model for predicting $\hx_{t+1 \mid t}$ and $\P_{t+1 \mid t}$: 
    \begin{subequations}
        \begin{align}
            \hx_{t+1 \mid t} & \overset{(\ref{eq-ssm-transition})}{=} \A \cdot \hx_{t \mid t} , \label{eq-predictor-prediction-xt}\\  
            \P_{t+1 \mid t} & \overset{(\ref{eq-ssm-transition})}{=}\A \cdot \P_{t \mid t}\cdot \A^H+\B \cdot \bSigma_{\u} \cdot \B^H. \label{eq-predictor-prediction-cov}
        \end{align}	
        \label{eq-predictor-prediction}
    \end{subequations}

    The predicted signal is obtained as:
    \begin{subequations}
        \begin{align}
            \hy_{t+1 \mid t} & \overset{(\ref{eq-ssm-observation})}{=} \D \cdot \hx_{t+1 \mid t},\\  
            \S_{t+1 \mid t} &\overset{(\ref{eq-ssm-observation})}{=}\D \cdot \P_{t+1 \mid t}\cdot \D^H+\bSigma_{\v} . \label{eq-predictor-expect-cov}
        \end{align}	
    \end{subequations}

    The predicted CSI vector $\hh_{t+1 \mid t}$ can then be recovered from the first $MN$ elements of $\hx_{t+1 \mid t}$ as:
    \begin{equation}
        \hh_{t+1 \mid t} = \B^{\top} \cdot \hx_{t+1 \mid t}.
        \label{eq-extract-h}
    \end{equation}
\end{enumerate}

\begin{remark}  \textbf{(Model Mismatch)}   
    As shown in (\ref{eq-predictor-filtering-kt}), the conventional Kalman gain $\mathbf{K}_t$ essentially functions as a weighting matrix, fusing the predicted information from $\hat{\mathbf{x}}_{t \mid t-1}$  and the observed information from $\Delta \mathbf{y}_t$\cite{gustafsson2010statistical}. However, (\ref{eq-kg}) reveals that the validity of $\mathbf{K}_t$ depends on the covariance terms $\mathbf{P}_{t \mid t-1}$ and $\mathbf{S}_{t \mid t-1}$, which are influenced by the approximated linear transition model (mismatched with the actual highly nonlinear channel dynamics), as evidenced in (\ref{eq-predictor-prediction-cov}) and (\ref{eq-predictor-expect-cov}). This model mismatch compromises the weighting capacity of $\mathbf{K}_t$ \cite{revach2022kalmannet}, and degrades the filtering performance.
 \end{remark}

\begin{remark} \textbf{(Objective Misalignment)} 
    Kalman gain was originally derived to minimize the filtering error $\|\x_{t}-\hx_{t \mid t}\|^2$ \cite{sarkka2023bayesian}, or equivalently, $\|\h_{t}-\hh_{t \mid t}\|^2$ in the context of channel prediction. This just serves as an intermediate step before minimizing the prediction error $\|\x_{t+1}-\hx_{t+1 \mid t}\|^2$, or equivalently, $\|\h_{t+1}-\hh_{t+1 \mid t}\|^2$ (\ref{eq-objective-predict}) in this context. This objective misalignment makes it difficult for $\mathbf{K}_t$ to effectively handle the performance loss in the prediction process (\ref{eq-predictor-prediction-xt}) due to the mismatched transition model.
 \end{remark}

 These issues arising from limited expert knowledge echo the necessity of a novel weighting matrix to replace the deviated Kalman gain in the FTP workflow. This novel weighting matrix should possess sufficient expressiveness to compensate for model mismatch \cite{revach2022kalmannet}, while remaining aligned with the prediction objective. Such a novel weighting matrix can be learned from data by neural networks, which forms the basis of our proposed hybrid FTP workflow in Section \ref{sec-method-kpin}.

\subsection{KPIN-enabled Hybrid FTP Workflow}
\label{sec-method-kpin}
\noindent This subsection introduces a principled integration of the data-driven neural network into the model-based FTP workflow, resulting in the KPIN-enabled hybrid FTP workflow, to address the question \textbf{Q1} raised in Section \ref{sec-intro}.

\begin{figure}[!t]
  \centering
  \includegraphics[width=\columnwidth]{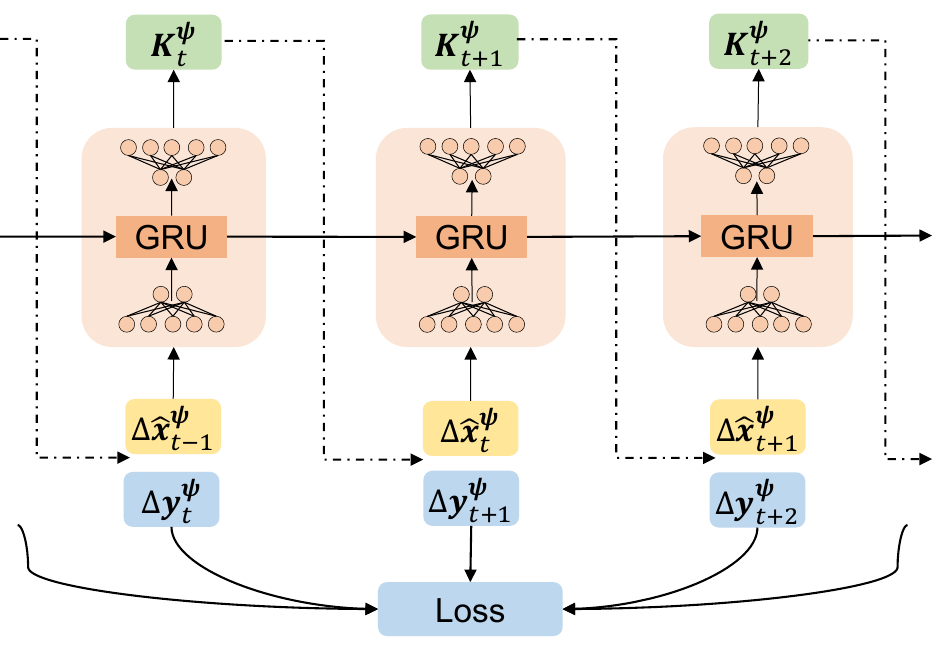}
  \caption{Hybrid FTP workflow consists of multiple KPIN modules with shared parameters $\bpsi$, depicted in the orange areas, invoked at each time step. Solid arrows connect data-driven components (\ref{eq-output-kpin}). Dashed arrows represent model-based components (\ref{eq-kpin-posterior}, \ref{eq-kpin-input}, \ref{eq-kpin-predict}, \ref{eq-kpin-extract}), as further illustrated in Fig. \ref{fig-train}. Blue rectangles correspond to the loss terms in (\ref{eq-ls}).}
  \label{fig-kpin}
\end{figure}

Fig. \ref{fig-kpin} illustrates the information flows within the KPIN-enabled hybrid FTP workflow. This workflow follows a recursive process of filtering and prediction over time, akin to conventional methods. The key distinction is the integration of a data-driven weighting matrix, $\mathbf{K}_t^{\bpsi} \in \mathbb{C}^{pMN \times \tau N}$, estimated by KPIN $g_{\bpsi}(\cdot)$—a neural network parameterized by $\bpsi$. For clarity, terms influenced by KPIN are denoted with a special superscript $\{\cdot\}^{\bpsi}$.

\begin{enumerate}
    \item \textbf{Filter:}
    
    Based on the prior estimate $\hx_{t \mid t-1}^{\bpsi}$ and the predicted signal $\hy_{t \mid t-1}^{\bpsi}$, a KF-type filtering process is implemented to yield the posterior estimate $\hx_{t \mid t}^{\bpsi}$ \cite{revach2022kalmannet}:
    \begin{equation}
        \hx_{t \mid t}^{\bpsi} = \hx_{t \mid t-1}^{\bpsi}+ \K_t^{\bpsi} \cdot \Delta \y_t^{\bpsi} ,
        \label{eq-kpin-posterior}
    \end{equation}
    where $\K_t^{\bpsi}$ is learned directly from data:
            \begin{subequations}
              \begin{align}
                \k_t^{\bpsi}&=\vec(\K_t^{\bpsi}) \in \bbC^{pMN^2\tau}, \\  
                  \left[
                \begin{array}{c}
                \k_t^{\bpsi, \Re}\\
                \k_t^{\bpsi, \Im}
                \end{array}
            \right] 
                &= g_{\bpsi}(\left[
                    \begin{array}{l}
                    \Delta \y_t^{\bpsi,\Re}\\
                    \Delta \y_t^{\bpsi,\Im}\\
                    \Delta \hx_{t-1}^{\bpsi,\Re}\\
                    \Delta \hx_{t-1}^{\bpsi,\Im}
                    \end{array}
                \right] ),
            \label{eq-output-kpin}
              \end{align}	
          \end{subequations}

        where the input features are defined as:
        \begin{subequations}
            \begin{align}
                \Delta \y_{t}^{\bpsi} & = \y_{t}-\hy_{t \mid t-1}^{\bpsi},  \label{eq-delta-y} \\  
                \Delta \hx_{t-1}^{\bpsi} & = \hx_{t-1 \mid t-1}^{\bpsi}-\hx_{t-1 \mid t-2}^{\bpsi}.
            \end{align}	
            \label{eq-kpin-input}
        \end{subequations}

        \item \textbf{Predict:}
        
        Next, $\hx_{t \mid t}^{\bpsi}$ passes through the SSM for obtaining the prior estimate $\hx_{t+1 \mid t}^{\bpsi}$ and the predicted signal $\hy_{t+1 \mid t}^{\bpsi}$:
        \begin{subequations}
            \begin{align}
                \hx_{t+1 \mid t}^{\bpsi} & \overset{(\ref{eq-ssm-transition})}{=} \A \cdot \hx_{t \mid t}^{\bpsi} , \label{eq-kpin-prior} \\  
                \hy_{t+1 \mid t}^{\bpsi} & \overset{(\ref{eq-ssm-observation})}{=} \D \cdot \hx_{t+1 \mid t}^{\bpsi}.
            \label{eq-kpin-expect}
            \end{align}	
            \label{eq-kpin-predict}
        \end{subequations}
        
        Our goal, the predicted CSI vector $\hh_{t+1 \mid t}^{\bpsi}$, can then be recovered from the first $MN$ elements of $\hx_{t+1 \mid t}$:
        \begin{equation}
            \hh_{t+1 \mid t}^{\bpsi}  \ = \B^{\top} \cdot \hx_{t+1 \mid t}^{\bpsi}
            \label{eq-kpin-extract}
        \end{equation}
\end{enumerate}

It is important to note that the complex channel dynamics can be implicitly tracked by the internal structure of KPIN. KPIN includes a gated recurrent unit (GRU) \cite{chung2014empirical} positioned between two fully connected (FC) layers. The hidden state of the GRU, different from the latent state of the SSM, is updated and propagated over time to capture the temporal patterns governing the generation of $\mathbf{K}_t^{\bpsi}$, as depicted by the solid arrows between GRUs in Fig. \ref{fig-kpin}. If the hidden state is not updated, i.e., the solid arrows are cut off, the GRU module essentially degrades into a typical MLP, significantly impairing ability of KPIN to track fast-changing channels in the long term. This critical role of the GRU hidden state update is further validated by the ablation simulations shown in Fig. \ref{fig-nmse_GRU_ablation} in Section \ref{sec-result-ablation}.

\subsection{Low-complexity Unsupervised Learning}
\label{sec-method-train}
\noindent All previous data-driven channel prediction methods \cite{kim2020massive, zhu2019adaptive, wei2022channel, jiang2022accurate} train neural networks iteratively using gradient descent (GD) to optimize specific objective functions \cite{goodfellow2016deep}. This often requires a significant amount of labeled CSI data, which is rarely available in practical scenarios. 

In contrast, KPIN is trained by maximizing the following conditional likelihood of $\left\{\mathbf{y}_t\right\}_{t=1}^{T}$, conditioned on the prior estimates influenced by $\bpsi$. This introduces a novel unsupervised learning strategy \cite{gama2023unsupervised,revach2022unsupervised}, requiring only a single sequence of historically received noisy signals, without the need for costly manual labeling of $\left\{\mathbf{h}_t\right\}_{t=1}^{T}$. This effectively addresses \textbf{Q2} raised in Section \ref{sec-intro}:
\begin{equation}
    \argmax_{\bpsi} \ p( \y_1,\ldots, \y_T \mid  \hx_{1 \mid 0}^{\bpsi}, \ldots, \hx_{T \mid T-1}^{\bpsi}),
    \label{eq-ml}
\end{equation}
which is equivalent to minimizing the following negative log-likelihood function based on Proposition \ref{pro-ML}:
\begin{equation}
    \argmin_{\bpsi} \ \sum_{t=0}^{T-1} \|\underbrace{\y_{t+1}-\Q \cdot \bPhi \cdot (\hx_{t \mid t-1}^{\bpsi}+\K_t^{\bpsi} \cdot \Delta \y_t^{\bpsi})}_{\Delta \y_{t+1}^{\bpsi}}\|^2.
    \label{eq-ls}
\end{equation}

\begin{proposition}
    Under the linear Gaussian SSM (\ref{eq-ssm-aug}) and the hybrid FTP workflow (\ref{eq-kpin-posterior}-\ref{eq-kpin-extract}), the optimization problems in (\ref{eq-ml}) and (\ref{eq-ls}) share the same solution set. 
    \label{pro-ML}
\end{proposition}

\begin{italicproof}
    \noindent \tb{Proof.} Mainly by transforming the predicted CSI vector into the predicted signal vector, detailed as follows:
    \begin{subequations}
        \begin{align}
             & \argmax_{\bpsi} p\left( \y_1,\ldots, \y_T \mid  \hx_{1 \mid 0}^{\bpsi}, \ldots, \hx_{T \mid T-1}^{\bpsi} \right) \label{eq-ml-1}\\ 
            = \  &  \argmax_{\bpsi} \prod_{t=0}^{T-1} p\left( \y_{t+1} \mid  \hx_{t+1 \mid t}^{\bpsi} \right) \label{eq-ml-3}\\
            \overset{(\ref{eq-ssm-observation})}{=} &  \argmin_{\bpsi} \sum_{t=0}^{T-1} \left(\y_{t+1}-\D \cdot \hx_{t+1 \mid t}^{\bpsi}\right)^H \cdot \bSigma_{\v}^{-1} \notag \\
            & \qquad \qquad \quad \cdot \left(\y_{t+1}-\D \cdot \hx_{t+1 \mid t}^{\bpsi}\right) \\  
            = \  &   \argmin_{\bpsi} \sigma_{\v}^{-2} \cdot \sum_{t=0}^{T-1} \left\|\y_{t+1}-\D \cdot \hx_{t+1 \mid t}^{\bpsi}\right\|^2 \\
            = \  &  \argmin_{\bpsi} \sum_{t=0}^{T-1} \left\|\y_{t+1}-\D \cdot \hx_{t+1 \mid t}^{\bpsi}\right\|^2 \\
            \overset{(\ref{eq-ssm-transition})}{=}  & \argmin_{\bpsi} \sum_{t=0}^{T-1} \left\|\y_{t+1}-\Q \cdot \bPhi \cdot \hx_{t \mid t}^{\bpsi}\right\|^2 \\
            \overset{(\ref{eq-kpin-posterior})}{=} & \argmin_{\bpsi} \sum_{t=0}^{T-1} \left\|\y_{t+1}-\Q \cdot \bPhi \cdot (\hx_{t \mid t-1}^{\bpsi}+\K_t^{\bpsi} \cdot \Delta \y_t^{\bpsi})\right\|^2,
            \label{eq-gnet-loss-expanded}
        \end{align}
        \label{eq-label}
    \end{subequations}

        \noindent where the transition to (\ref{eq-ml-3}) holds due to the i.i.d. assumption in the signal model (\ref{eq-transmission-vec}). 

        \hfill $\blacksquare$
\end{italicproof}

Training KPIN directly by solving (\ref{eq-ls}) leads to two drawbacks. Firstly, it incurs a computational burden of $\BigO(T^2)$, scaling quadratically with the length $T$ of the entire training sequence. This is because back-propagating the error term $\|\Delta \mathbf{y}_{t+1}^{\bpsi}\|^2$ in (\ref{eq-ls}) requires calling the KPIN module $t$ times due to the recursive structure illustrated in Fig. \ref{fig-kpin}. Secondly, empirical results indicate that this method leads to a less stable training process. Hence, the single sequence is segmented into $n_s \in \bbN^+$ short subsequences with a length of $T_{s} \in \bbN^+$, $T_s \ll T$, resulting in a segmented training dataset $\{ \{\mathbf{y}_{t,j}\}_{t=1}^{T_{s}}\}_{j=1}^{n_s}$.

\begin{algorithm}[t]
    \caption{Unsupervised learning with noisy signals}
     \begin{algorithmic} 
        \State \textbf{Inputs:} Historically received noisy signals $\{ \y_t\}_{t=1}^{T}$
    \end{algorithmic}
    \begin{algorithmic}[1] 
            \State Segment $\{ \y_t\}_{t=1}^{T}$ into $\{ \left\{\y_{t,j}\right\}_{t=1}^{T_{s}}\}_{j=1}^{n_s}$
            \State Initialize KPIN parameters $\bpsi^{(0)}$ 
            \For{$i=1$ to $n_{e}$}    
                \For{$j=1$ to $n_{b}$}
                    \State Randomly select index $b \in \{n_s\}$ 
                    \State Initialize $\hx_{0 \mid 0}$ for subsequence $\left\{\y_{t,b}\right\}_{t=1}^{T_{s}}$
                    \For{$t=0$ to $T_s-1$}
                        \State Filter posterior estimate $\hx_{t \mid t,j}^{\bpsi}$ as (\ref{eq-kpin-posterior})
                        \State Predict prior estimate $\hx_{t+1 \mid t,j}^{\bpsi}$ as (\ref{eq-kpin-prior})
                        \State Predict received signal $\hy_{t+1 \mid t,j}^{\bpsi}$ as (\ref{eq-kpin-expect})
                        \State Compute loss $\ell_{j} (\bpsi^{(i-1)}, t+1)$ as (\ref{eq-gnet-single-step})
                    \EndFor
                \EndFor
                \State Compute epoch objective $\mathcal{L}(\bpsi^{(i-1)})$ as (\ref{eq-gnet-batchloss})
                \State Back-propagate gradients $\nabla_{\bpsi} \mathcal{L}(\bpsi^{(i-1)})$
                \State Update KPIN parameters $\bpsi^{(i-1)} \rightarrow \bpsi^{(i)}$
            \EndFor
    \end{algorithmic}
    \begin{algorithmic} 
        \State \textbf{Outputs:} Optimal KPIN parameters $\bpsi^* \approx \bpsi^{(n_{e})}$
    \end{algorithmic}
    \label{alg-gnet}
\end{algorithm}

In line with data segmentation, during each training epoch, $n_b \in \bbN^+$ subsequences ($n_b \leq n_{s}$) are randomly selected to compute the following objective function in each epoch \cite{revach2022kalmannet}. This results in a mini-batch GD method with performance guarantees \cite{goodfellow2016deep}, while reducing the total training expenses to only $\BigO(n_b\cdot T_s^2)$ (Numerical validation is provided in Fig.~\ref{fig-nmse_vs_nb} and the corresponding comments in Section \ref{sec-result-scenario}.):

\begin{equation}
    \begin{aligned}
        \mathcal{L}(\bpsi)
        \triangleq \frac{1}{n_b\cdot T_{s}} \sum_{j=1}^{n_b} \sum_{t=0}^{T_s-1}\ell_{j} (\bpsi, t+1)+\beta \cdot\|\bpsi\|,
    \end{aligned}
    \label{eq-gnet-batchloss}
  \end{equation}

  \noindent where $\beta \in \bbR^+$ is a regularization factor to avoid overfitting. The single-step loss $\ell_{j} (\bpsi, t+1)$ for the  $(t+1)$-th step of the $j$-th subsequence is defined based on (\ref{eq-ls}):
\begin{equation}
  \begin{aligned}
    \ell_{j} (\bpsi, t+1)
      \triangleq \|\underbrace{\y_{t+1,j}-\Q \cdot \bPhi \cdot (\hx_{t \mid t-1,j}^{\bpsi}+\K_{t,j}^{\bpsi} \cdot \Delta \y_{t,j}^{\bpsi})}_{\Delta \y_{t+1,j}^{\bpsi}}\|^2.
  \end{aligned}
  \label{eq-gnet-single-step}
\end{equation}

\noindent The training process lasts $n_e \in \bbN^+$ epochs, as outlined in Algorithm \ref{alg-gnet}.

\subsection{Discussions}
\label{sec-method-discuss}
\noindent This subsection dissects the hybrid FTP workflow and the unsupervised learning strategy to interpret the benefits arising from the hybrid integration of KPIN, thereby addressing \textbf{Q3} in Section \ref{sec-intro}.

KPIN is interpretable primarily because it leverages neural networks only for the most complex part of the hybrid FTP workflow, specifically the computation of $\K_t^{\bpsi}$ (\ref{eq-output-kpin}), as shown in Fig. \ref{fig-kpin}. On one hand, compared to purely data-driven methods that directly learn channel dynamics without incorporating physical knowledge, this hybrid method preserves the main structure and, consequently, theoretical foundation of the FTP workflow, resulting in a principled integration. On the other hand, the model-based components in the hybrid FTP workflow possess explicit physical meanings, enhancing its interpretability—an advantage often overlooked in conventional data-driven methods, which are typically treated as "black boxes" in practice.

\begin{figure}[!t]
    \centering
    \includegraphics[width=\columnwidth]{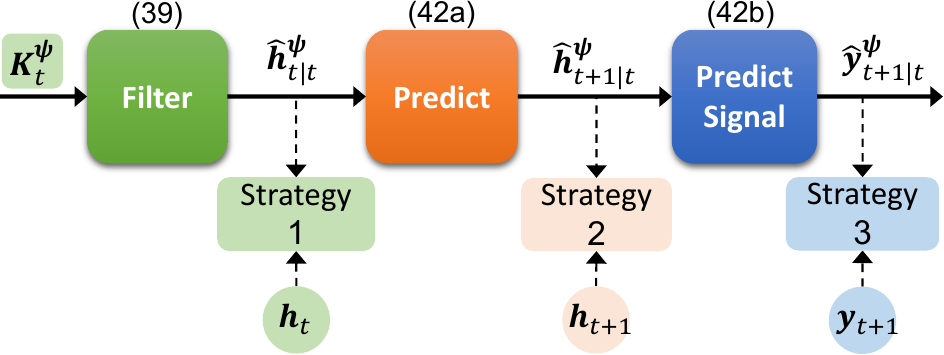}
    \caption{Information flow through the model-based components. }
    \label{fig-train}
\end{figure}

\begin{table}[t]
    \caption{Comparison of three supervision strategies}
    \centering
    \renewcommand{\arraystretch}{2.3} 
    \begin{tabularx}{\columnwidth}{>{\centering\arraybackslash}p{1.1cm}>{\centering\arraybackslash}X*{2}{>{\centering\arraybackslash}p{2.2cm}}}
        \hline 
        \textbf{Strategy} & \textbf{1} & \textbf{2} & \textbf{3} (KPIN)\\
        \hline 
        \hline
        \textbf{Supervision} & $\h_t$&  $\h_{t+1}$ & $\y_{t+1}$\\
        \textbf{Objective} & $\|\h_t-\hh_{t \mid t}^{\bpsi}\|^2$&$\|\h_{t+1}-\hh_{t+1 \mid t}^{\bpsi}\|^2$&$\|\y_{t+1}-\hy_{t+1 \mid t}^{\bpsi}\|^2$ \\
        \textbf{Matched} & \textcolor{blue}{\ding{52}}&\textcolor{blue}{\ding{52}}& \textcolor{blue}{\ding{52}}\\
        \textbf{Aligned} &\textcolor{red}{\ding{56}}&\textcolor{blue}{\ding{52}}&\textcolor{blue}{\ding{52}}\\
        \textbf{Unlabeled}  &\textcolor{red}{\ding{56}}&\textcolor{red}{\ding{56}}&\textcolor{blue}{\ding{52}}\\
        \hline  
    \end{tabularx}
    \vspace{5pt}
    
    \raggedright

    \textbf{Matched:} model mismatch corrected in filtering. \textbf{Aligned:} aligned with the prediction objective. \textbf{Unlabeled:} supervised by unlabeled data.
    \label{tab-supervision}
\end{table}

More specifically, the model-based components can be divided into three consecutive steps, enabling us to examine its internal information flows for deeper insights \footnote{There is a slight abuse of notation in this subsection. The CSI vector serves as our target in the context of channel prediction. Thus, for brevity, we denote $\hh_{t \mid t}^{\bpsi}$ and $\hh_{t+1 \mid t}^{\bpsi}$ instead of $\hx_{t \mid t}^{\bpsi}$ and $\hx_{t+1 \mid t}^{\bpsi}$, respectively, since the former can be recovered from the initial $MN$ elements of the latter, similar to (\ref{eq-kpin-extract}).}: 
\begin{equation}
    \Delta \y_{t+1}^{\bpsi} \overset{(\ref{eq-ls})}{=} \y_{t+1} -  \underbrace{ \Q \cdot \underbrace{\bPhi \cdot \underbrace{(\hx_{t \mid t-1}^{\bpsi}+\K_t^{\bpsi} \cdot \Delta \y_t^{\bpsi})}_{\text{\ding{172} Filter } \rightarrow \ \hh_{t \mid t}}}_{\text{\ding{173} Predict } \rightarrow \ \hh_{t+1 \mid t}}}_{\text{\ding{174} Predict Signal }\rightarrow \ \hy_{t+1 \mid t}}
    \label{eq-gnet-factorized}
\end{equation}

To further illustrate these concepts, Fig. \ref{fig-train} visualizes (\ref{eq-gnet-factorized}), showcasing how the intermediate information flows are utilized to construct three distinct objective functions, corresponding to the supervision strategies outlined in Table \ref{tab-supervision}. This setup facilitates an ablation analysis to better understand the progressively enhanced properties from Strategy 1 to Strategy 3, ultimately elucidating the benefits of our hybrid method. 

Further details are discussed below:

\begin{enumerate}
    \item \textbf{Strategy 1} supervises the filtered posterior estimate $\hh_{t \mid t}^{\bpsi}$ to minimize the filtering error $\|\h_t-\hh_{t \mid t}^{\bpsi}\|^2$, an intermediate step before minimizing the prediction error $\|\h_{t+1}-\hh_{t+1 \mid t}^{\bpsi}\|^2$. This strategy originates from KalmanNet \cite{revach2022kalmannet}, with an extension from the real to the complex values made by us, resulting in what we call ComplexKalmanNet. It generates a data-driven weighting matrix to overcome model mismatch during filtering, thereby yielding a more accurate posterior estimate. However, since its objective function remains misaligned with the objective of channel prediction (\ref{eq-objective-predict}), the improved posterior estimate may still be distorted by the mismatched transition model in the consequent prediction step (\ref{eq-predictor-prediction-xt}), leading to inaccurate prediction. Additionally, Strategy 1 necessitates extensive labeled CSI data, typically obtained through LMMSE \cite{tse2005fundamentals}, which involves significant human effort and may introduce unwanted pre-processing errors. These limitations motivate the development of more effective strategies that directly optimize prediction errors without relying on labeled data, leading to Strategies 2 and 3.

    \item \textbf{Strategy 2} supervises the predicted CSI vector $\hh_{t+1 \mid t}^{\bpsi}$ to directly minimize prediction errors, by further integrating the prediction step into its objective function compared to Strategy 1. This adaptation effectively addresses both model mismatch and objective misalignment by accounting for potential errors at every stage of the FTP workflow. Consequently, it introduces a novel weighting matrix $\K_t^{\bpsi}$ that significantly alters the functionality of the conventional Kalman gain $\K_t$, resulting in a substantially improved prediction accuracy. However, it still relies on labor-intensive manual labeling to acquire high-quality labels for $\{\h_{t+1}\}_{t=0}^{T-1}$. Can the manual labeling process be automated? This question motivates the development of Strategy 3 (KPIN).

    \item \textbf{Strategy 3.} The signal model (\ref{eq-transmission-vec}) provides valuable physical insights by transforming the predicted CSI $\hh_{t+1 \mid t}^{\bpsi}$ into the predicted signal $\hy_{t+1 \mid t}^{\bpsi}$. The truly received signal $\y_{t+1}$ is a realization drawn from a conditional distribution, concretely conditioned on $\Q \cdot \h_{t+1}$ with zero-mean AWGN, denoted as $\bbE [\y_{t+1} \mid \h_{t+1}] \overset{(\ref{eq-transmission-vec})}{=}\Q \cdot \h_{t+1}$. Hence, from a statistical perspective, $\y_{t+1}$ effectively serves as labeled data for the predicted signal $\hy_{t+1 \mid t}^{\bpsi}\overset{(\ref{eq-transmission-vec})}{=}\Q \cdot\hh_{t+1 \mid t}^{\bpsi}$. Furthermore, the transformed pilot matrix $\Q$ has full column rank, eliminating any non-identifiability issue \cite{frigola2014variational} in the signal model. This ensures that $\hh_{t+1 \mid t}^{\bpsi}$ can be uniquely recovered (or identified) from $\y_{t+1}^{\bpsi}$. These factors altogether lead to the favorable result that minimizing $\|\y_{t+1}-\hy_{t+1 \mid t}^{\bpsi}\|^2$ is statistically equivalent to optimizing $\|\h_{t+1}-\hh_{t+1 \mid t}^{\bpsi}\|^2$ in Strategy 2. Strategy 3 also brings two other significant advantages over Strategies 1 and 2. Firstly, manually labeling $\h_{t+1}$ is implicitly encoded as an intermediate step of optimizing $\|\y_{t+1}-\hy_{t+1 \mid t}^{\bpsi}\|^2$, removing the need for labeled CSI data, thus lowering deployment barriers of KPIN in real-world scenarios. Secondly, this implicit labeling process reduces the risk of pre-processing errors that can occur during manual labeling.
\end{enumerate}

The progressive improvement from Strategy 1 to Strategy 3 (KPIN) is further supported by the ablation simulations depicted in Fig. \ref{fig-nmse_ablation} of Section \ref{sec-result-ablation}, which demonstrate a gradual decrease in prediction errors.

\section{Numerical Results}
\label{sec-result}

\begin{algorithm}[t]
    \caption{Testing KPIN on futurely received signals}
     \begin{algorithmic} 
        \State \textbf{Inputs:} $\{\y_t^{\text{test}}\}_{t=1}^{L}$$\triangleq$ futurely received signals $\{ \y_t\}_{t=T+1}^{T+L}$
    \end{algorithmic}
    \begin{algorithmic}[1] 
            \State Initialize KPIN parameters $\bpsi=\bpsi^*$ 
                    \State Initialize $\hx_{0 \mid 0}$ 
                    \For{$t=0$ to $L-1$}
                    \State Filter posterior estimate $\hx_{t \mid t}^{\bpsi^*}$ as (\ref{eq-kpin-posterior})
                    \State Predict prior estimate $\hx_{t+1 \mid t}^{\bpsi^*}$ as (\ref{eq-kpin-prior})
                    \State Predict received signal $\hy_{t+1 \mid t}^{\bpsi^*}$ as (\ref{eq-kpin-expect})
                    \State Extract predicted CSI $\hh_{t+1 \mid t}^{\bpsi^*}$ as (\ref{eq-kpin-extract})
            \EndFor
    \end{algorithmic}
    \begin{algorithmic} 
        \State \textbf{Outputs:} Predicted CSI $\{ \hh_{t+1 \mid t}\}_{t=T}^{T+L-1}$$\triangleq$$\{ \hh_{t+1 \mid t}^{\bpsi^*}\}_{t=0}^{L-1}$
    \end{algorithmic}
    \label{alg-gnet-test}
  \end{algorithm}

\subsection{Experimental Settings}
\label{sec-result-set}
\noindent The QuaDRiGa (QUAsi Deterministic RadIo channel GenerAtor) simulator \cite{jaeckel2014quadriga1,jaeckel2014quadriga2} is utilized for data synthesis owing to its compatibility with the 3GPP mmWave channel model, \textit{3GPP\_TR\_38.901} \cite{3gpp_TS_38.901}. This simulator can capture the intricate short-term time evolution of channel parameters, delivering an authentic portrayal of time-varying channels impacted by UE movement in realistic and complicated environments. In our implementation, we use QuaDRiGa to simulate the channels between the UE and BS based on the dynamic locations of the UE at different time steps $t$, effectively capturing the time-varying nature of the channels as experienced by the moving UE.

 In pursuit of practical relevance, the carrier frequency $f$ is selected to align with the frequency bands tentatively reserved for forthcoming commercial deployment, as reported by Qualcomm \cite{qualcomm}. The parameter $\rho$ is tuned to attain the desired SNR values. We set $M=\tau$ and the pilot matrix as follows for simplicity \cite{marzetta2016fundamentals}:
 \begin{equation}
    \Q_{\text{pilot}} = \sqrt{\tau} \cdot \I_M. 
    \label{eq-pilot-define}
\end{equation}
The trained KPIN is tested using Algorithm \ref{alg-gnet-test} over a segment of futurely received noisy signals $\{ \y_t\}_{t=T+1}^{T+L}$. The accompanying source code is publicly available online. \footnote{\href{https://github.com/sunyiyong/KPIN}{https://github.com/sunyiyong/KPIN}}

The predictive performance is evaluated using the normalized squared error (NSE) for each future time step $(t+1)$, defined as:
\begin{equation}
    \begin{aligned} 
        \text{NSE}_{t+1} \triangleq & \frac{\left\|\h_{t+1}-\hh_{t+1 \mid t}\right\|^2}{\left\|\h_{t+1}\right\|^2},
    \end{aligned}
    \label{eq-nse}
\end{equation}
where $\h_{t+1}$ is the ground truth CSI vector for the future time step $(t+1)$, and $\hh_{t+1 \mid t}$ is the predicted CSI vector given the received signals up to the time step $t$.

The channel prediction task (\ref{eq-objective-predict}) treats all episodes of the consecutive $(p+1)$ time steps as a realization of the same stochastic process \cite{haykin2002adaptive}. Therefore, to assess the overall predictive performance over a future horizon, the normalized mean squared error (NMSE) is obtained by averaging the NSEs over $L$ future steps as:
\begin{equation}
    \begin{aligned} 
        \text{NMSE} \triangleq  \bbE \left[\text{NSE}_{t+1}\right]\approx  \frac{1}{L}\sum_{t=T}^{T+L-1} \text{NSE}_{t+1}.
    \end{aligned}
    \label{eq-nmse}
\end{equation}

The performance of KPIN is compared against state-of-the-art channel prediction methods, including two purely model-based approaches (AR \cite{kim2020massive} and ARKF \cite{kim2020massive} that is based on the conventional FTP workflow) and two purely data-driven methods (GRU \cite{wei2022channel} and Transformer \cite{jiang2022accurate}).  

\begin{remark}
    It is important to note that KPIN operates without access to the ground truth CSI data during the training and testing phases, which reflects a more realistic scenario, putting it at a disadvantage compared to its competitors. For instance, AR, GRU and Transformer utilize ${\h_{t-p},\ldots, \h_t}$ as inputs, while both GRU and Transformer are supervised by $\h_{t+1}$.
\end{remark}

All methods are implemented using PyCharm 2022.2.4 (Professional Edition) on a system equipped with an Intel(R) Xeon(R) Gold 6230R CPU @ 2.10GHz processor and a Quadro RTX 6000 GPU. The implementation utilizes Python 3.8.5, PyTorch 1.7.1, and CUDA 11.0 for efficient GPU acceleration. The Adam optimizer \cite{kingma2014adam} is employed for training neural networks.

\begin{figure}[t]
  \centering
  \includegraphics[width=\columnwidth]{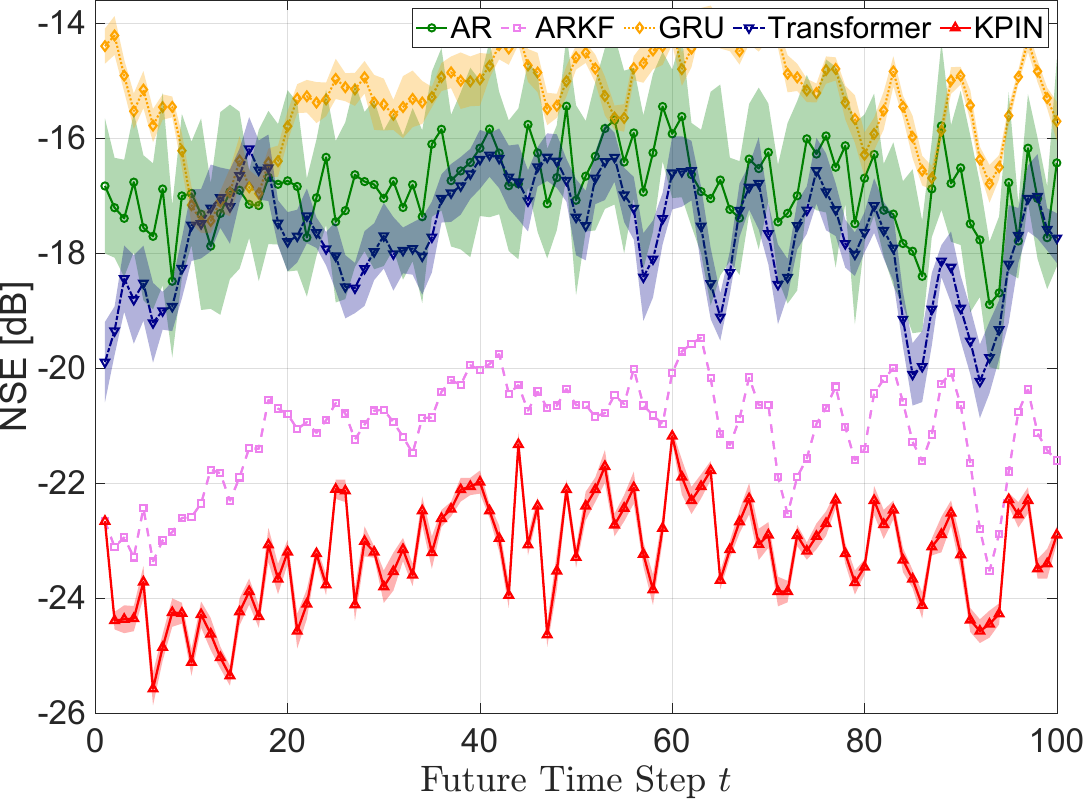}
  \caption{Mean and standard deviation of NSE versus future time step for five methods. }
  \label{fig-nmse_qualitative}
\end{figure}

\subsection{Qualitative Analysis}
\label{sec-result-quali}
\noindent The performance evaluation begins with a qualitative comparison of five methods in Fig. \ref{fig-nmse_qualitative} under baseline conditions, while a comprehensive analysis across various scenario settings is provided in Section \ref{sec-result-scenario} for quantitative comparison. The baseline conditions includes: BS antenna height of 25 meters, UE antenna height of 1.5 meters, $N=32$, $M =2$, $\tau=2$, $v=60$ km/h, $f=28$ GHz, $k=\frac{\Delta t}{T_c}=1$, SNR$=20$ dB, $p=4$, $T=1000$, $L=100$, $T_s=10$, $n_b=50$, and $n_e=100$. The learning rate is set as $5 \times 10^{-5}$ while $\beta = 1 \times 10^{-5}$. The results are averaged over ten Monte Carlo runs with different random seeds. The performance of ARKF is not affected by random seeds, hence no standard deviation is reported. The high speed of $v=60$ km/h introduces substantial Doppler effects, leading to rapid changes in the channel dynamics. In line with realistic scenarios, all methods rely on a training length of only $T=1000$ to capture this intricate temporal pattern for predicting the CSI in the future $L=100$ steps.

\begin{figure}[t]
    \centering 
    \includegraphics[width=\columnwidth]{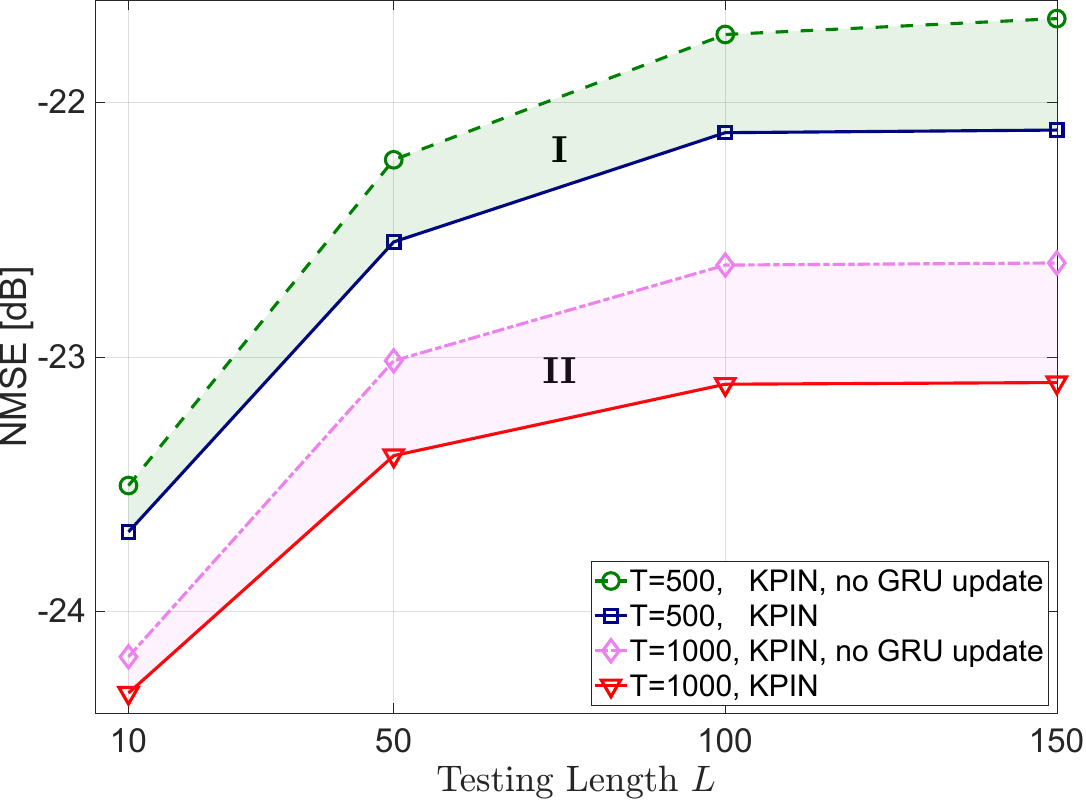}
    \caption{Comparison results of different internal structures of KPIN obtained from an ablation study.}
    \label{fig-nmse_GRU_ablation}
  \end{figure}

Fig. \ref{fig-nmse_qualitative} illustrates that the proposed KPIN consistently achieves the lowest prediction error throughout the entire prediction horizon, with an improvement of approximately -2 dB over ARKF, which achieves the second-best performance. The similarity in their trends indicates that KPIN can effectively learn to implement the FTP workflow similar to the conventional model-based one. Their performance gap emphasizes the advantages of the hybrid FTP workflow, utilizing expressive neural networks to overcome model mismatch and objective misalignment inherent in the conventional model-based FTP workflow. This will be further explained in Section \ref{sec-result-ablation}.

Despite the ground truth CSI data $\{\h_{t-p},\ldots, \h_t\}$ as inputs, AR struggles with its linear structure, which is unsuitable to model the highly nonlinear channel dynamics. Lacking the FTP workflow, it showcases the largest standard deviation of performance among all methods. Though supervised by the ground truth CSI data $\h_{t+1}$, both data-driven GRU and Transformer exhibit significantly larger prediction errors of around -15 dB and -18 dB, respectively. This can be attributed to their lack of incorporation of physical knowledge, highlighting the efficacy of methods like ARKF and KPIN based on the principled FTP workflow. The advantage of Transformer over GRU lies in its attention mechanism \cite{vaswani2017attention}, which captures more contextual information about time-varying channels, compared to the simple sequential architecture of GRU.

\subsection{Ablation Studies}  
\label{sec-result-ablation}
\noindent Two ablation studies are conducted to investigate the observed performance gains of KPIN in the qualitative analysis. These empirical results provide concrete evidence of the capability of KPIN to implicitly track channel dynamics, as discussed at the end of Section \ref{sec-method-kpin}, and highlight the mutual benefits of KPIN mentioned in Section \ref{sec-method-discuss}.

Fig. \ref{fig-nmse_GRU_ablation} demonstrates the critical importance of the GRU hidden state update in enhancing the long-term prediction performance of KPIN. With a training length of $T=500$, the KPIN without the GRU hidden state update shows a higher NMSE compared to the KPIN with the update. This performance gap, highlighted by the colored Region I, consistently increases at a diminishing rate as the prediction horizon extends from $L=10$ to $L=150$. This observation aligns with our previous statement in Section \ref{sec-method-kpin} that a GRU without updates essentially functions as an MLP, lacking the capacity to adapt to fast-changing channel dynamics over the long term. When the training length is doubled from $T=500$ to $T=1000$, the performance improvement due to the GRU hidden state update becomes even more pronounced, as shown by comparing the colored Region II to Region I. This enhancement occurs because the additional training data allows KPIN to more accurately learn the channel dynamics. The aforementioned results provide valuable insights for future designs of more powerful KPIN structures (e.g., based on the attention mechanism in Transformer \cite{vaswani2017attention}) aimed at improving prediction accuracy.                                                                                                                                                     

\begin{figure}[t]
    \centering 
    \includegraphics[width=\columnwidth]{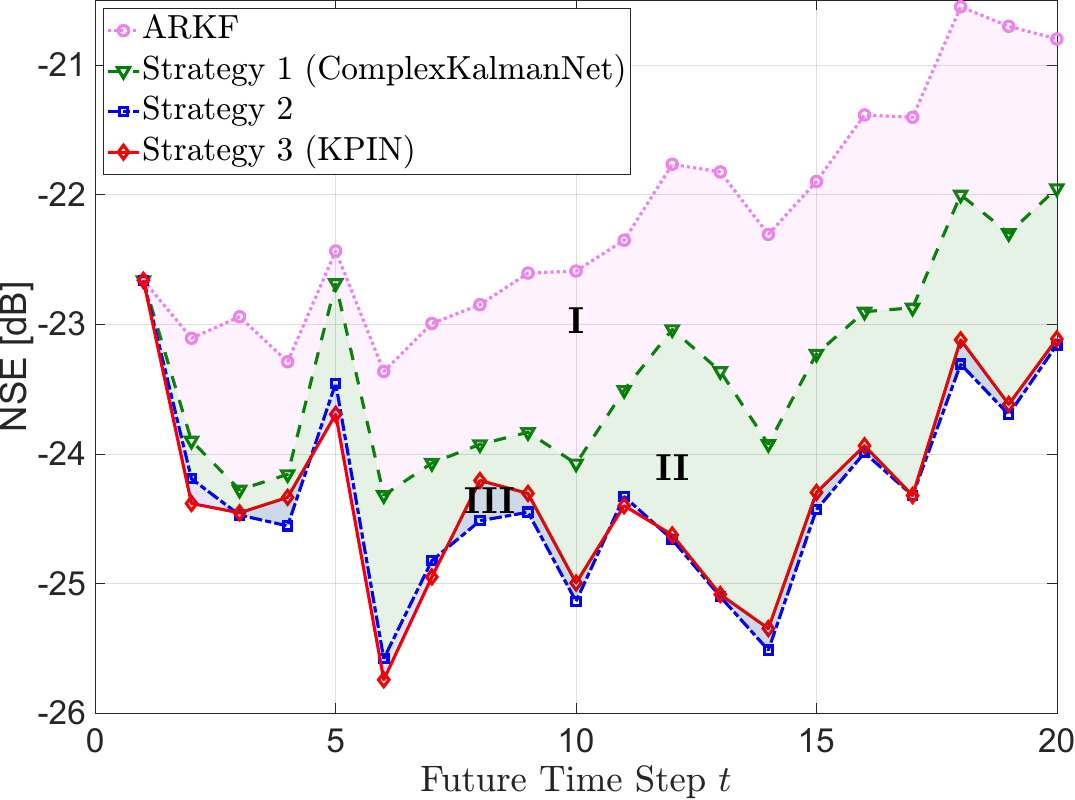}
    \caption{Comparison results of different supervision strategies obtained from an ablation study.}
    \label{fig-nmse_ablation}
  \end{figure}

In Fig. \ref{fig-nmse_ablation}, the three strategies exhibit a consistent trend in the prediction performance compared to ARKF, indicating that they have effectively learned and applied the FTP workflow. Their performance gaps are color-coded, representing the incremental benefits from Strategy 1 (ComplexKalmanNet) to Strategy 3 (KPIN). Such improvement comes from step-by-step addressing model mismatch, objective misalignment, and avoiding the need for manual labeling, as outlined in Table~\ref{tab-supervision}. Concretely, we have the following observations from different regions in Fig. \ref{fig-nmse_ablation}, supporting the claims made in Section \ref{sec-method-discuss}.

\begin{figure}[t]
    \centering 
    \includegraphics[width=0.98\columnwidth]{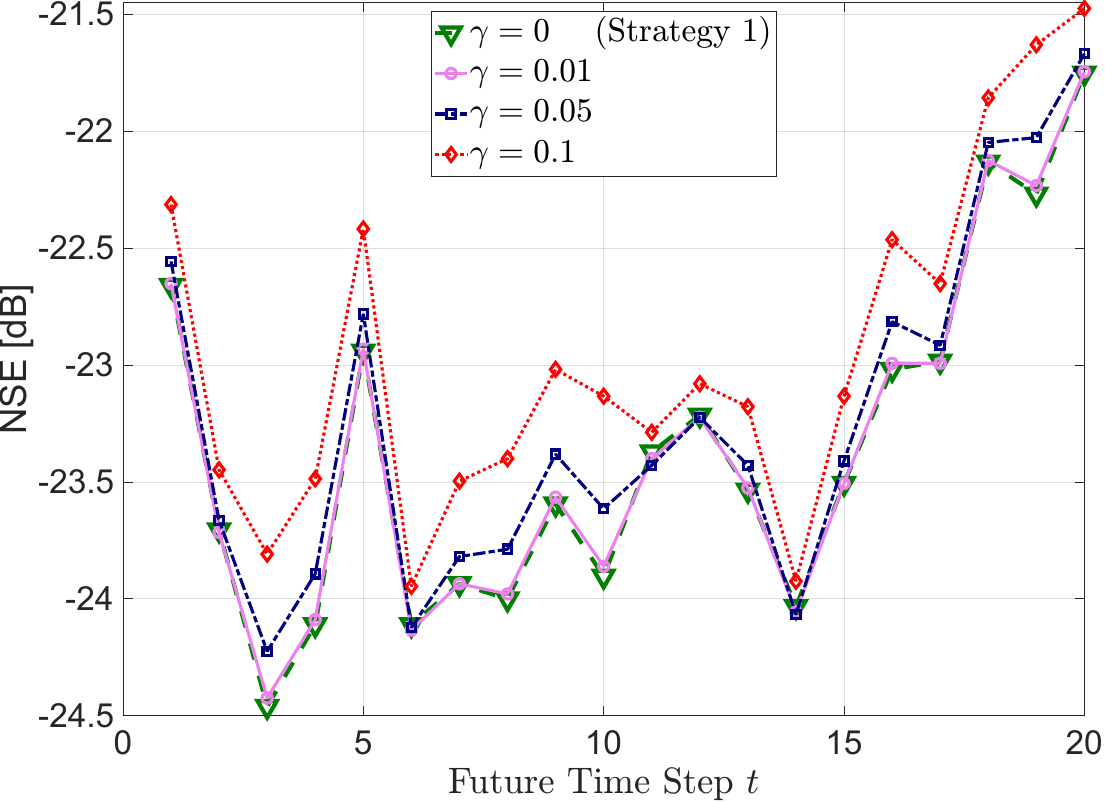}
     \caption{Comparison of prediction performance with different levels of pre-processing errors added to the labels in Strategy 1. The resulting supervision strategy is denoted as $\|\h_t + \gamma \cdot \mathbf{n}_t - \hh_{t \mid t}^{\bpsi}\|^2$, where the pre-processing error $\mathbf{n}_t \in \bbC^{MN}$ is i.i.d. AWGN with a similar magnitude to $\h_t$ and $\gamma \in \bbR^+$ controls the error level.}
    \label{fig-posterior_h_noise}
  \end{figure}

  \begin{figure}[t]
    \centering \includegraphics[width=0.98\columnwidth]{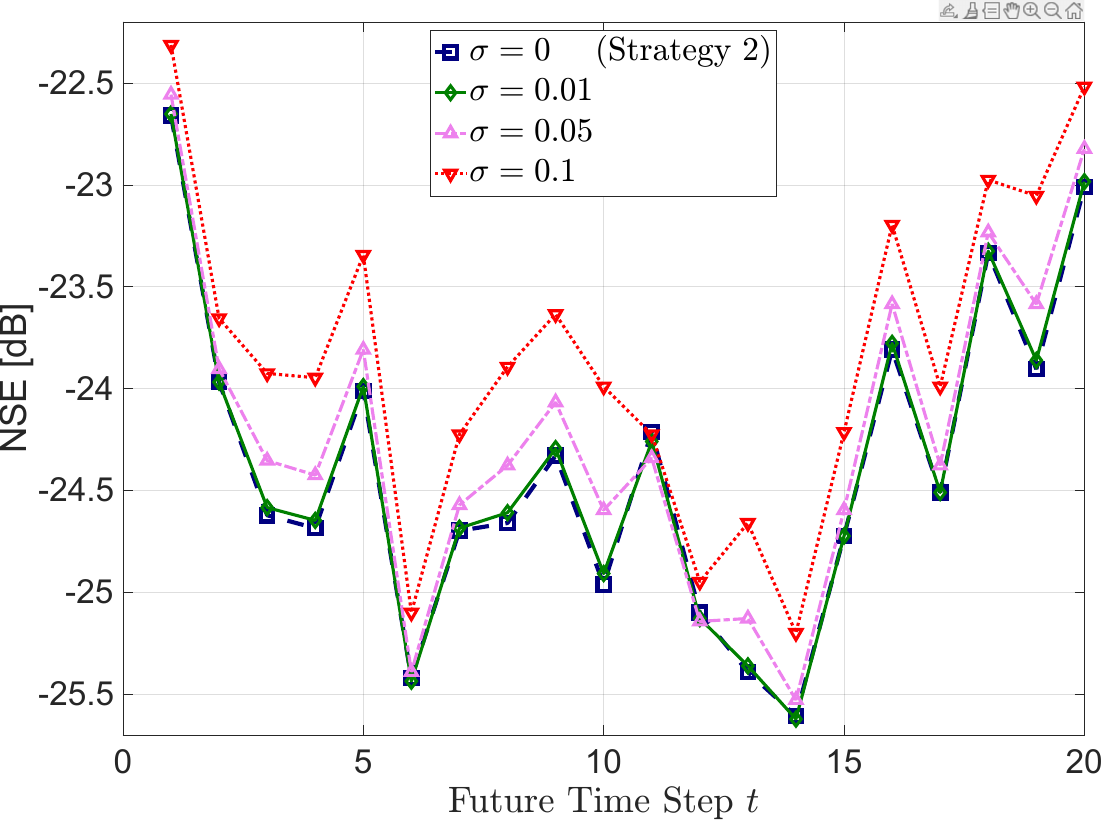}
    \caption{Comparison of prediction performance with different levels of pre-processing errors added to the labels in Strategy 2. The resulting supervision strategy is denoted as $\|\h_{t+1} + \sigma \cdot \mathbf{n}_{t+1} - \hh_{t+1 \mid t}^{\bpsi}\|^2$, where the pre-processing error $\mathbf{n}_{t+1}\in \bbC^{MN}$ is i.i.d. AWGN with a similar magnitude to $\h_{t+1}$ and $\sigma \in \bbR^+$ controls the error level.}
    \label{fig-prior_h_noise}
  \end{figure}

\textbf{Region I:} Strategy 1 demonstrates an error reduction of over -1 dB in most time steps compared to ARKF. This improvement is attributed to its integration of a data-driven weighting matrix to optimize the filtering error $\|\h_t-\hh_{t \mid t}^{\bpsi}\|^2$. While ARKF suffers from the deviated Kalman gain based on the mismatched SSM (\ref{eq-ssm-aug}), Strategy 1 effectively addresses model mismatch by leveraging an expressive neural network in its hybrid FTP workflow, leading to an increased prediction precision.

\textbf{Region II:} Moving to Strategy 2, a further performance gain of about -1 dB can be observed in Region II. Aligned with the optimization objective $\|\h_{t+1}-\hh_{t+1 \mid t}^{\bpsi}\|^2$ of the channel prediction task, Strategy 2 introduces a novel weighting matrix $\K_t^{\bpsi}$ that simultaneously addresses both model mismatch and objective misalignment. In total, it results in a substantial decrease of about 2 dB in the prediction error compared to that of ARKF.

\textbf{Region III:} The comparable performance of Strategies 2 and 3 (KPIN) empirically proves the statistical equivalence between minimizing $\|\y_{t+1}-\hy_{t+1 \mid t}^{\bpsi}\|^2$ and optimizing $\|\h_{t+1}-\hh_{t+1 \mid t}^{\bpsi}\|^2$, as claimed in Section \ref{sec-method-discuss}. Their performance difference is negligible, especially when considering the significant benefits of KPIN in eliminating the need for manual labeling. It is important to note that in real-world applications, Strategies 1 and 2 may experience higher prediction errors, as they rely solely on the manual labels of $\h_{t+1}$ for supervision, which can introduce pre-processing errors compared to the ground truth. This phenomenon is illustrated in Fig. \ref{fig-posterior_h_noise}, where the constant $\gamma$ varies from 0.01 to 0.1, indicating an increase in pre-processing errors for Strategy 1, which leads to a higher prediction error for Strategy 1. A similar trend is evident in Fig. \ref{fig-prior_h_noise}, where the prediction error for Strategy 2 increases with larger pre-processing errors, as controlled by an increasing $\sigma$.

\begin{figure}[t]
  \centering
  \includegraphics[width=\columnwidth]{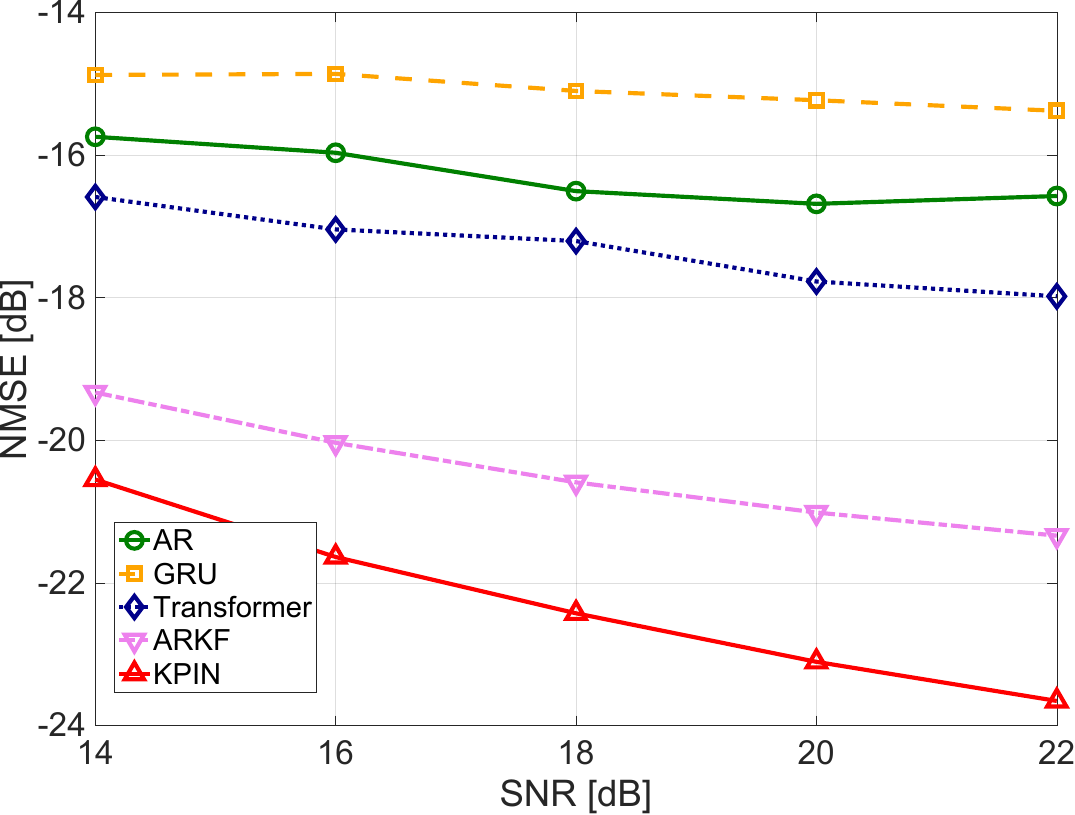}
  \caption{NMSE versus SNR for five methods.}
  \label{fig-nmse_vs_SNR}
\end{figure}

\subsection{Robustness towards Different Scenario Configurations}
\label{sec-result-scenario}
\noindent Considering the practical deployment, it is crucial to evaluate the effectiveness of KPIN under diverse challenging conditions, including low SNR, exacerbated channel aging and the increasing number of MIMO antennas in future wireless networks. These analyses provide valuable insights into the robustness and adaptability of KPIN in real-world applications.

Fig. \ref{fig-nmse_vs_SNR} indicates a decrease in the prediction errors for all methods as SNR increases. Consistent with our previous observations in Fig. \ref{fig-nmse_qualitative}, the performance gains of ARKF and KPIN compared to the other three methods are evident, demonstrating the effectiveness of the principled FTP workflow in reducing errors. Specifically, in the most challenging scenario with an SNR of only 14 dB, both ARKF and KPIN achieve a prediction error of approximately -20 dB, a significant improvement over the error of -18 dB observed for Transformer in a more favorable scenario with an SNR of 22 dB. Of particular interest is that KPIN consistently maintains a significant performance improvement over ARKF, even at lower SNR levels. This reaffirms the inherent ability of KPIN to overcome model mismatch and objective misalignment present in ARKF, unaffected by varying noise levels. 

\begin{figure}[t]
    \centering
    \includegraphics[width=\columnwidth]{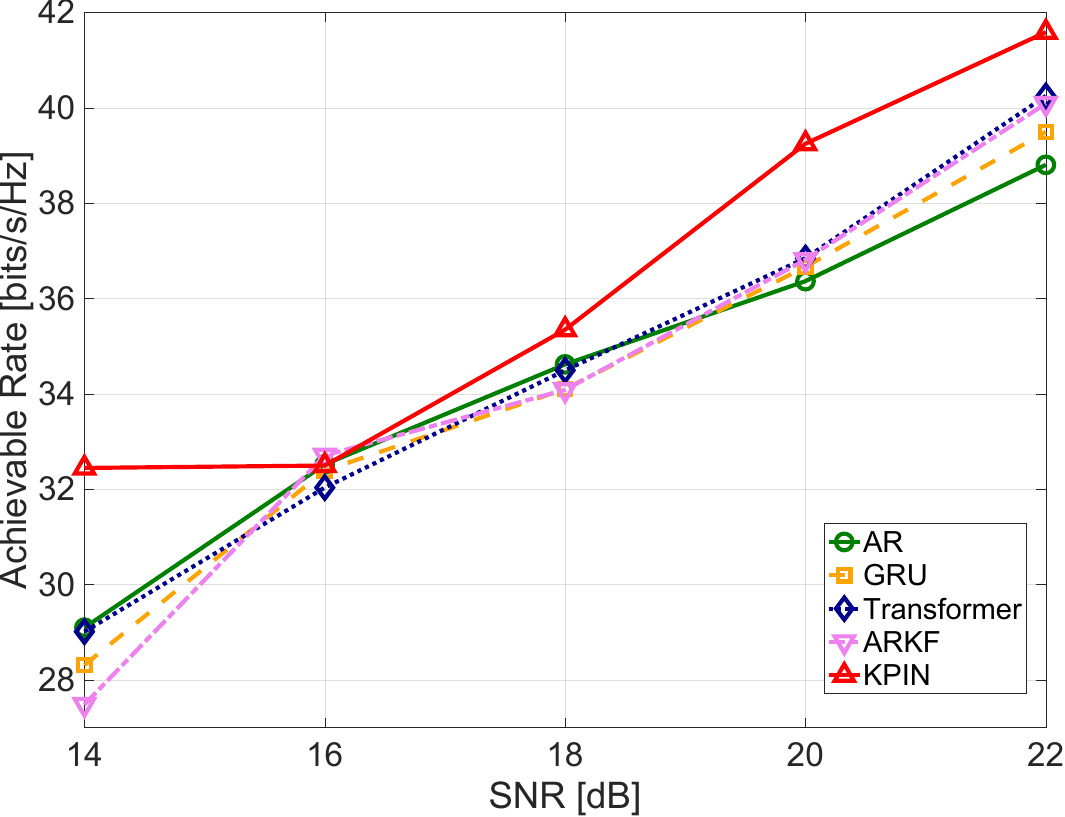}
    \caption{Achievable rate versus SNR for five methods.}
    \label{fig-rate_vs_SNR}
\end{figure}

\begin{figure}[t]
    \centering
    \includegraphics[width=\columnwidth]{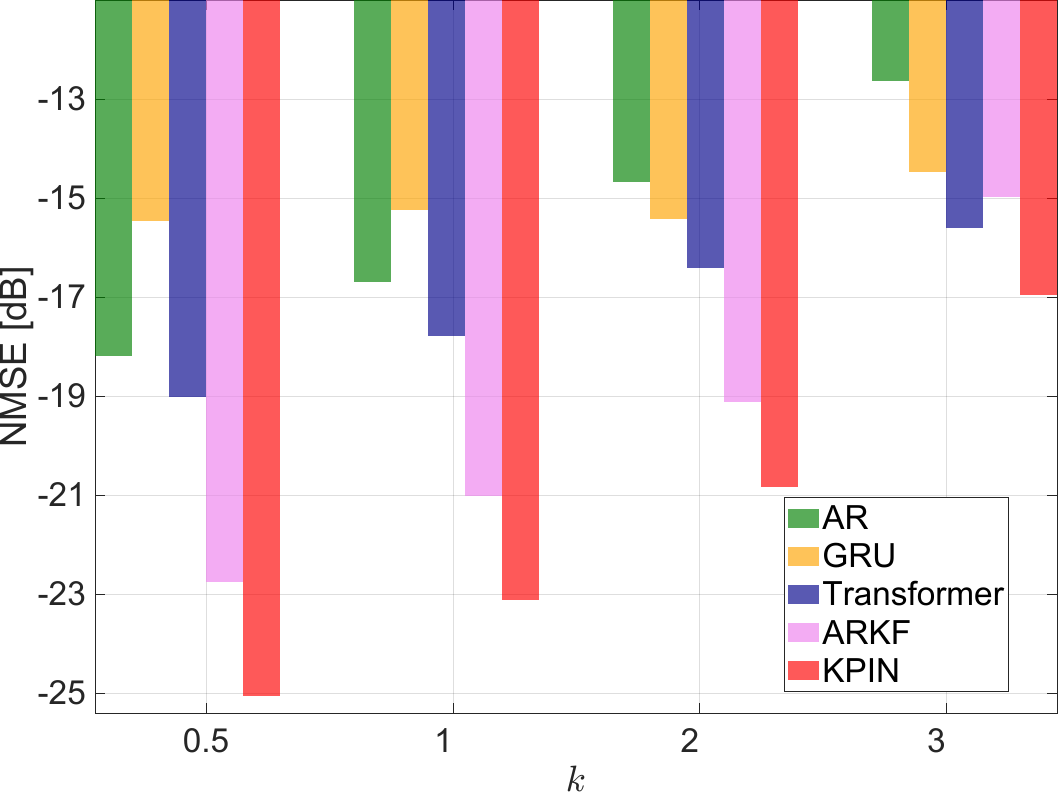}
    \caption{NMSE versus severity of channel aging (indicated by $k$) for the five methods.}
    \label{fig-nmse_vs_delta_t}
\end{figure}

In addition, Fig.~\ref{fig-rate_vs_SNR} compares the achievable rate, defined in (\ref{eq-rate}), for the five methods. As the SNR increases from 14 dB to 22 dB, the achievable rate for all methods shows significant growth. The proposed KPIN outperforms its competitors, particularly demonstrating an advantage of about 3 bits/s/Hz in a challenging scenario with an SNR of only 14 dB, while consistently maintaining almost the highest achievable rate as SNR increases. These observations are consistent with both the theoretical analyses and ablation simulation results, further underscoring the effectiveness of hybrid methods in tracking time-varying channels compared to purely model-based or data-driven methods.

Fig. \ref{fig-nmse_vs_delta_t} demonstrates the prediction performance of five methods across increasing levels of channel aging. Remark~\ref{remark-k} illustrates that a larger $k$ indicates more severe channel aging, leading to higher prediction errors across all methods. KPIN consistently exhibits the lowest prediction errors for various $k$ values. This advantage arises from its expressive internal structure with the GRU hidden state update, which enables KPIN to effectively and implicitly track fast-changing channels.

\renewcommand{\arraystretch}{2.2}
\begin{table}[t] 
    \caption{NMSE versus BS antenna number $N$}
    \centering
    \begin{tabularx}{\columnwidth}{>{\bfseries}c*{5}{>{\centering\arraybackslash}X}}
    \hline    
    & \textbf{16} & \textbf{32} & \textbf{48} & \textbf{64} & \textbf{128}\\
    \hline
    \hline
    \textbf{AR} & -17.26 & -16.68 & -16.36 & -16.27 & -15.81 \\
    \textbf{GRU} & -16.63 & -15.23 & -14.09 & -13.23 & -13.16 \\
    \textbf{Transformer} & -18.44 & -17.77 & -17.43 & -17.24 & -16.45 \\
    \textbf{ARKF} & -21.37 & -21.01 & -21.28 & -21.33 & -21.27 \\
     \textbf{KPIN} & -23.00 & -23.11 & -23.86 & -24.12 & -24.87 \\
    \hline
    \end{tabularx}
    \label{tab-bs-antenna}
    \raggedright
\end{table}

Table~\ref{tab-bs-antenna} illustrates the prediction NMSE with respect to the number of BS antennas. An increase in $N$ augments the dimensionality of channels and signals, posing challenges for the data-driven GRU and Transformer. Lacking reasonable incorporation of physical knowledge, these methods necessitate heavily overparameterized neural networks to effectively capture the dynamics of time-varying channels. This incurs a substantially exacerbated demand for training data \cite{bach2024learning}, a condition often impractical to fulfill. In contrast, ARKF maintains a lower NMSE thanks to its model-based FTP workflow, which exhibits relative insensitivity to changes in dimensionality. Facilitated by its hybrid FTP workflow, KPIN inherits this robustness towards variation in antenna numbers, showcasing a promising trend for handling future systems with hundreds of antennas.

\begin{figure}[t]
    \centering
    \includegraphics[width=\columnwidth]{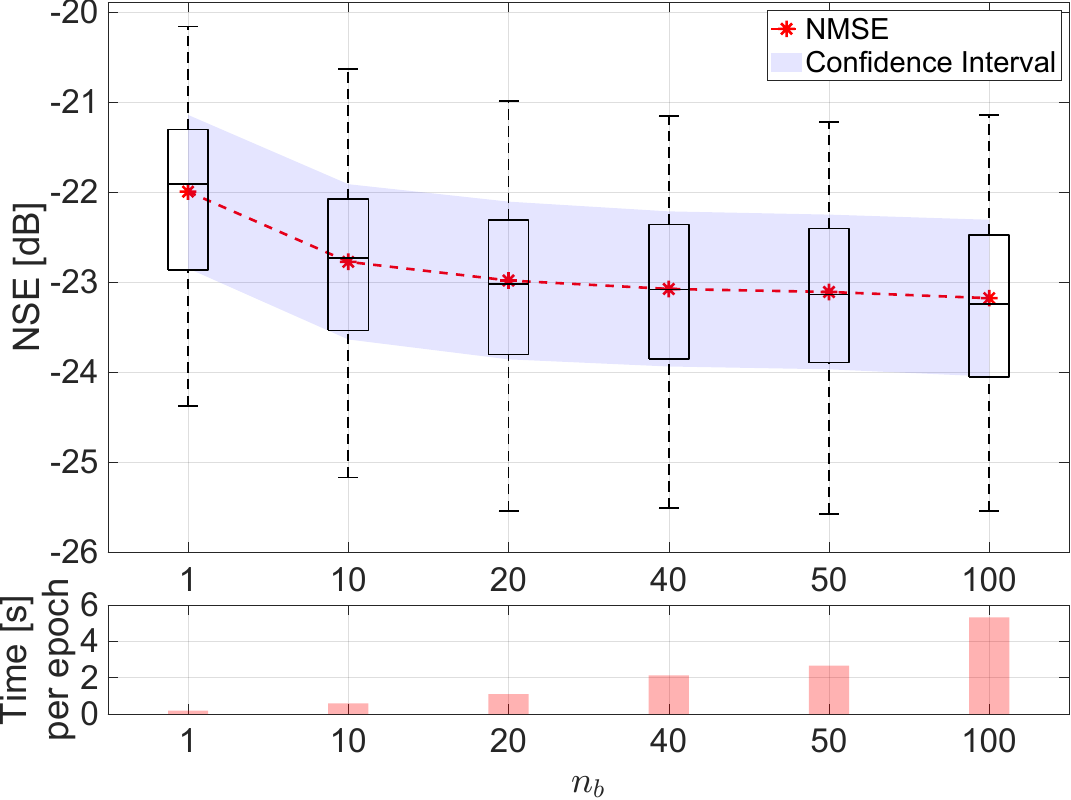}
    \caption{NMSE and epoch time versus batch size for KPIN.}
    \label{fig-nmse_vs_nb}
  \end{figure}

Fig. \ref{fig-nmse_vs_nb} provides insights into selecting the batch size $n_b$, which controls the trade-off between the prediction performance and training expenses of KPIN. The total training length $T=1000$ and the subsequence length $T_s=10$ results in $n_s=100$ subsequences. The upper part of Fig. \ref{fig-nmse_vs_nb} illustrates that as $n_b$ increases from 1 to 100 (i.e., from stochastic GD to mini-batch GD to full-batch GD \cite{goodfellow2016deep}), the prediction error decreases constantly till $n_b=20$ and then remains stable thereafter. The bottom part of Fig. \ref{fig-nmse_vs_nb} demonstrates that the training time per epoch increases linearly with $n_b$, aligning with the training complexity of $\BigO(n_b \cdot T_s^2)$ as was discussed in Section \ref{sec-method-train}. These observations heuristically suggest an optimal batch size of $n_b=20$, which ensures the best trade-off. It reduces the total training complexity from $\BigO(1000^2)$ for training over the entire sequence to only $\BigO(2000)$ for training over segmented subsequences. This further confirms the effectiveness of the proposed strategies of data segmentation and mini-batch GD.

\renewcommand{\arraystretch}{2.2}
\begin{table}[t]  
    \caption{NMSE versus training length $T$}
    \centering
    \begin{tabularx}{\columnwidth}{c*{4}{>{\centering\arraybackslash}X}}
    \hline
     & \textbf{500} & \textbf{1000} & \textbf{1500} & \textbf{2000}\\
    \hline
    \hline
    \textbf{AR} & -16.08 & -16.68 & -17.04 & -17.16 \\
    \textbf{GRU} & -16.68 & -15.23 & -15.06 & -15.06 \\
    \textbf{Transformer} & -15.97 & -17.77 & -18.54 & -18.81 \\
    \textbf{ARKF} & -20.74 & -21.01 & -20.88 & -20.54 \\
    \textbf{KPIN} & -22.18 & -23.11 & -23.04 & -22.80 \\
    \hline
    \end{tabularx}
    \label{tab-train-length}
    \raggedright
\end{table}

Table~\ref{tab-train-length} summarizes the prediction performance across various training sequence lengths for different prediction methods. It is observed that as the training length is increased by four times, namely from $T=500$ to $T=2000$, Transformer exhibits a notable improvement in prediction accuracy, with an NMSE drop of almost 3 dB. In contrast to the data-hungry nature of Transformer, ARKF utilizes physical knowledge from the SSM to mitigate the dependence on training data volume. This enables ARKF to achieve a substantially superior NMSE of -20.74 dB with a modest training sequence length of only $T=500$. The low data demand observed in ARKF is also evident in KPIN, which adopts a hybrid FTP workflow. KPIN achieves a further reduction in NMSE by effectively addressing issues such as model mismatch and objective misalignment encountered in ARKF. The combination of data efficiency and enhanced prediction accuracy underscores the distinct advantage of KPIN, particularly in real-world deployment with limited availability of training data.

\section{Conclusion}
\label{sec-conclusion}
\noindent This paper introduces a novel high-accuracy method for predicting highly time-varying channels under large user mobility with limited expert knowledge. The proposed KPIN method integrates classical model-based and data-driven methods, effectively addressing issues arising from limited expert knowledge to enhance prediction accuracy. An innovative unsupervised learning strategy is developed to eliminate the need for manual labeling, significantly reducing barriers to practical deployment. The mutual benefits of KPIN are theoretically analyzed and empirically validated through ablation studies. Numerical experiments demonstrate substantial performance gain of KPIN over state-of-the-art channel prediction methods, both in terms of prediction errors and achievable rates. Comprehensive evaluations assess the robustness and adaptability of KPIN against various challenging factors, positioning it as a promising candidate for next-generation wireless communication, particularly in scenarios involving high user mobility, low SNR, and large antenna arrays.

\bibliographystyle{IEEEtran}
\bibliography{reference}

\begin{thebibliography}{10}
\providecommand{\url}[1]{#1}
\csname url@samestyle\endcsname
\providecommand{\newblock}{\relax}
\providecommand{\bibinfo}[2]{#2}
\providecommand{\BIBentrySTDinterwordspacing}{\spaceskip=0pt\relax}
\providecommand{\BIBentryALTinterwordstretchfactor}{4}
\providecommand{\BIBentryALTinterwordspacing}{\spaceskip=\fontdimen2\font plus
\BIBentryALTinterwordstretchfactor\fontdimen3\font minus
  \fontdimen4\font\relax}
\providecommand{\BIBforeignlanguage}[2]{{%
\expandafter\ifx\csname l@#1\endcsname\relax
\typeout{** WARNING: IEEEtran.bst: No hyphenation pattern has been}%
\typeout{** loaded for the language `#1'. Using the pattern for}%
\typeout{** the default language instead.}%
\else
\language=\csname l@#1\endcsname
\fi
#2}}
\providecommand{\BIBdecl}{\relax}
\BIBdecl

\bibitem{tmc_2023_v2x}
L.~Jiao, K.~Yu, J.~Chen, T.~Liu, H.~Zhou, and L.~Cai, ``Performance analysis of
  uplink/downlink decoupled access in cellular-v2x networks,'' \emph{IEEE
  Transactions on Mobile Computing}, vol.~23, no.~5, pp. 5616--5630, 2023.

\bibitem{tmc2021delay}
B.~N{\'e}meth, N.~Molner, J.~Mart{\'\i}n-P{\'e}rez, C.~J. Bernardos, A.~De~la
  Oliva, and B.~Sonkoly, ``Delay and reliability-constrained vnf placement on
  mobile and volatile 5g infrastructure,'' \emph{IEEE Transactions on Mobile
  Computing}, vol.~21, no.~9, pp. 3150--3162, 2021.

\bibitem{tmc_2022_air}
J.~Zhou, D.~Tian, Y.~Yan, X.~Duan, and X.~Shen, ``Joint optimization of
  mobility and reliability-guaranteed air-to-ground communication for uavs,''
  \emph{IEEE Transactions on Mobile Computing}, vol.~23, no.~1, pp. 566--580,
  2022.

\bibitem{tmc_2022_beamforming}
Y.~Chen, Y.~Huang, C.~Li, Y.~T. Hou, and W.~Lou, ``Turbo-hb: a sub-millisecond
  hybrid beamforming design for 5g mmwave systems,'' \emph{IEEE Transactions on
  Mobile Computing}, vol.~22, no.~7, pp. 4332--4346, 2022.

\bibitem{tmc_2020_beamforming}
P.~Paul, H.~Wu, and C.~Xin, ``Boost: A user association and scheduling
  framework for beamforming mmwave networks,'' \emph{IEEE Transactions on
  Mobile Computing}, vol.~20, no.~10, pp. 2924--2935, 2020.

\bibitem{gao2023metaloc}
J.~Gao, D.~Wu, F.~Yin, Q.~Kong, L.~Xu, and S.~Cui, ``{MetaLoc}: Learning to
  learn wireless localization,'' \emph{IEEE Journal on Selected Areas in
  Communications}, 2023.

\bibitem{he2024cramer}
J.~He, K.~C. Ho, W.~Xiong, H.~C. So, and Y.~J. Chun, ``{Cramér-Rao Lower
  Bound} analysis for elliptic localization with random sensor positions,''
  \emph{IEEE Transactions on Aerospace and Electronic Systems}, pp. 1--10,
  2024.

\bibitem{tmc2024sensing}
F.~Zhang, Z.~Zhang, L.~Kang, and A.~Zhou, ``mmtaa: A contact-less
  thoracoabdominal asynchrony measurement system based on mmwave sensing,''
  \emph{IEEE Transactions on Mobile Computing}, 2024.

\bibitem{he2023framework}
J.~He, F.~Yin, and H.~C. So, ``A framework for millimeter-wave multi-user
  {SLAM} and its low-cost realization,'' \emph{Signal Processing}, vol. 209,
  109018, 2023.

\bibitem{tmc2021robust}
A.~Pourkabirian and M.~H. Anisi, ``Robust channel estimation in multiuser
  downlink 5g systems under channel uncertainties,'' \emph{IEEE Transactions on
  Mobile Computing}, vol.~21, no.~12, pp. 4569--4582, 2021.

\bibitem{tmc2024efficient}
L.~Yu and T.~Ji, ``Efficient federated learning with channel status awareness
  and devices' personal touch,'' \emph{IEEE Transactions on Mobile Computing},
  2024.

\bibitem{jiang2019exploiting}
Z.~Jiang, S.~Chen, A.~F. Molisch, R.~Vannithamby, S.~Zhou, and Z.~Niu,
  ``Exploiting wireless channel state information structures beyond linear
  correlations: A deep learning approach,'' \emph{IEEE Communications
  Magazine}, vol.~57, no.~3, pp. 28--34, 2019.

\bibitem{tmc2020beam}
S.~K. Moorthy and Z.~Guan, ``Beam learning in mmwave/thz-band drone networks
  under in-flight mobility uncertainties,'' \emph{IEEE Transactions on Mobile
  Computing}, vol.~21, no.~6, pp. 1945--1957, 2020.

\bibitem{tmc2021context}
H.~Ding and K.~G. Shin, ``Context-aware beam tracking for 5g mmwave v2i
  communications,'' \emph{IEEE Transactions on Mobile Computing}, vol.~22,
  no.~6, pp. 3257--3269, 2021.

\bibitem{tmc2023online}
M.~Krunz, I.~Aykin, S.~Sarkar, and B.~Akgun, ``Online reinforcement learning
  for beam tracking and rate adaptation in millimeter-wave systems,''
  \emph{IEEE Transactions on Mobile Computing}, vol.~23, no.~2, pp. 1830--1845,
  2023.

\bibitem{he2023modeling}
J.~He, Y.~J. Chun, and H.~C. So, ``Modeling and performance analysis of
  blockchain-aided secure {TDOA} localization under random internet-of-vehicle
  networks,'' \emph{Signal Processing}, vol. 206, p. 108904, 2023.

\bibitem{chopra2021data}
R.~Chopra and C.~R. Murthy, ``Data aided {MSE}-optimal time varying channel
  tracking in massive {MIMO} systems,'' \emph{IEEE Transactions on Signal
  Processing}, vol.~69, pp. 4219--4233, 2021.

\bibitem{baddour2005autoregressive}
K.~E. Baddour and N.~C. Beaulieu, ``Autoregressive modeling for fading channel
  simulation,'' \emph{IEEE Transactions on Wireless Communications}, vol.~4,
  no.~4, pp. 1650--1662, 2005.

\bibitem{kashyap2017performance}
S.~Kashyap, C.~Moll{\'e}n, E.~Bj{\"o}rnson, and E.~G. Larsson, ``Performance
  analysis of {(TDD)} massive {MIMO} with {K}alman channel prediction,'' in
  \emph{Proceedings of IEEE International Conference on Acoustics, Speech and
  Signal Processing}, New Orleans, LA, USA, 2017, pp. 3554--3558.

\bibitem{zhu2019adaptive}
Y.~Zhu, X.~Dong, and T.~Lu, ``An adaptive and parameter-free recurrent neural
  structure for wireless channel prediction,'' \emph{IEEE Transactions on
  Communications}, vol.~67, no.~11, pp. 8086--8096, 2019.

\bibitem{kim2020massive}
H.~Kim, S.~Kim, H.~Lee, C.~Jang, Y.~Choi, and J.~Choi, ``Massive {MIMO} channel
  prediction: Kalman filtering vs. machine learning,'' \emph{IEEE Transactions
  on Communications}, vol.~69, no.~1, pp. 518--528, 2020.

\bibitem{wei2022channel}
Y.~Wei, M.-M. Zhao, A.~Liu, and M.-J. Zhao, ``Channel tracking and prediction
  for {IRS}-aided wireless communications,'' \emph{IEEE Transactions on
  Wireless Communications}, vol.~22, no.~1, pp. 563--579, 2022.

\bibitem{jiang2022accurate}
H.~Jiang, M.~Cui, D.~W.~K. Ng, and L.~Dai, ``Accurate channel prediction based
  on {Transformer}: Making mobility negligible,'' \emph{IEEE Journal on
  Selected Areas in Communications}, vol.~40, no.~9, pp. 2717--2732, 2022.

\bibitem{yin2020addressing}
H.~Yin, H.~Wang, Y.~Liu, and D.~Gesbert, ``Addressing the curse of mobility in
  massive {MIMO} with prony-based angular-delay domain channel predictions,''
  \emph{IEEE Journal on Selected Areas in Communications}, vol.~38, no.~12, pp.
  2903--2917, 2020.

\bibitem{wong2005joint}
I.~C. Wong and B.~L. Evans, ``Joint channel estimation and prediction for
  {OFDM} systems,'' in \emph{Proceedings of IEEE Global Telecommunications
  Conference}, vol.~4, St. Louis, MO, USA, 2005, pp. 5--pp.

\bibitem{kalman1960new}
R.~E. Kalman, ``A new approach to linear filtering and prediction problems,''
  \emph{Journal of Basic Engineering}, vol.~82, no.~1, pp. 35--45, 1960.

\bibitem{tse2005fundamentals}
D.~Tse and P.~Viswanath, \emph{Fundamentals of Wireless Communication}.\hskip
  1em plus 0.5em minus 0.4em\relax Cambridge University Press, 2005.

\bibitem{revach2022kalmannet}
G.~Revach, N.~Shlezinger, X.~Ni, A.~L. Escoriza, R.~J. Van~Sloun, and Y.~C.
  Eldar, ``{KalmanNet}: Neural network aided {K}alman filtering for partially
  known dynamics,'' \emph{IEEE Transactions on Signal Processing}, vol.~70, pp.
  1532--1547, 2022.

\bibitem{sarkka2023bayesian}
S.~S{\"a}rkk{\"a} and L.~Svensson, \emph{Bayesian Filtering and
  Smoothing}.\hskip 1em plus 0.5em minus 0.4em\relax Cambridge University
  Press, 2023, vol.~17.

\bibitem{shlezinger2023model}
N.~Shlezinger, J.~Whang, Y.~C. Eldar, and A.~G. Dimakis, ``Model-based deep
  learning,'' \emph{Proceedings of the IEEE}, vol. 111, no.~5, pp. 465--499,
  2023.

\bibitem{3gpp_TS_38.901}
``{Study on channel model for frequencies from 0.5 to 100 GHz},'' 3rd
  Generation Partnership Project {(3GPP)}, TR 38.901 version 17.0.0, Mar. 2022.

\bibitem{goldsmith2005wireless}
A.~Goldsmith, \emph{Wireless Communications}.\hskip 1em plus 0.5em minus
  0.4em\relax Cambridge University Press, 2005.

\bibitem{liao2019ekf}
Y.~Liao, X.~Shen, G.~Sun, X.~Dai, and S.~Wan, ``Ekf/ukf-based channel
  estimation for robust and reliable communications in v2v and iiot,''
  \emph{EURASIP Journal on Wireless Communications and Networking}, vol. 2019,
  no.~1, p. 144, 2019.

\bibitem{tmc2018investigation}
P.~Linning, G.~Li, J.~Zhang, R.~Woods, M.~Liu, and A.~Hu, ``An investigation of
  using loop-back mechanism for channel reciprocity enhancement in secret key
  generation,'' \emph{IEEE transactions on mobile computing}, vol.~18, no.~3,
  pp. 507--519, 2018.

\bibitem{3gpp_TS_36.873}
``{Study on 3D channel model for LTE},'' 3rd Generation Partnership Project
  (3GPP), TR 36.873, Jan. 2018.

\bibitem{ITU-RM.2135}
``{Guidelines for evaluation of radio interface technologies for
  IMT-Advanced},'' International Telecommunication Union Radiocommunication
  Sector (ITU-R), M. 2135-1, Dec. 2009.

\bibitem{marzetta2016fundamentals}
T.~L. Marzetta and H.~Yang, \emph{Fundamentals of massive MIMO}.\hskip 1em plus
  0.5em minus 0.4em\relax Cambridge University Press, 2016.

\bibitem{haykin2002adaptive}
S.~S. Haykin, \emph{Adaptive filter theory}.\hskip 1em plus 0.5em minus
  0.4em\relax Pearson Education India, 2002.

\bibitem{yule1971method}
G.~U. Yule, ``On a method of investigating periodicities in disturbed series
  with special reference to {W}olfer’s sunspot numbers,'' \emph{Statistical
  Papers of George Udny Yule}, pp. 389--420, 1971.

\bibitem{walker1931periodicity}
G.~T. Walker, ``On periodicity in series of related terms,'' \emph{The Royal
  Society of London. Series A, Containing Papers of a Mathematical and Physical
  Character}, vol. 131, no. 818, pp. 518--532, 1931.

\bibitem{gustafsson2010statistical}
F.~Gustafsson, \emph{Statistical Sensor Fusion}.\hskip 1em plus 0.5em minus
  0.4em\relax Studentlitteratur, 2010.

\bibitem{chung2014empirical}
J.~Chung, C.~Gulcehre, K.~Cho, and Y.~Bengio, ``Empirical evaluation of gated
  recurrent neural networks on sequence modeling,'' in \emph{NeurIPS Workshop
  on Deep Learning, December}, 2014.

\bibitem{goodfellow2016deep}
I.~Goodfellow, Y.~Bengio, and A.~Courville, \emph{Deep Learning}.\hskip 1em
  plus 0.5em minus 0.4em\relax Cambridge, MA, USA: MIT Press, 2016.

\bibitem{gama2023unsupervised}
F.~Gama, N.~Zilberstein, M.~Sevilla, R.~G. Baraniuk, and S.~Segarra,
  ``Unsupervised learning of sampling distributions for particle filters,''
  \emph{IEEE Transactions on Signal Processing}, 2023.

\bibitem{revach2022unsupervised}
G.~Revach, N.~Shlezinger, T.~Locher, X.~Ni, R.~J. van Sloun, and Y.~C. Eldar,
  ``Unsupervised learned {K}alman filtering,'' in \emph{Proceedings of 30th
  European Signal Processing Conference}, Belgrade, Serbia, 2022, pp.
  1571--1575.

\bibitem{frigola2014variational}
R.~Frigola, Y.~Chen, and C.~E. Rasmussen, ``Variational {G}aussian process
  state-space models,'' \emph{Advances in Neural Information Processing
  Systems}, vol.~27, 2014.

\bibitem{jaeckel2014quadriga1}
S.~Jaeckel, L.~Raschkowski, K.~B{\"o}rner, and L.~Thiele, ``{QuaDRiGa}: A 3-{D}
  multi-cell channel model with time evolution for enabling virtual field
  trials,'' \emph{IEEE Transactions on Antennas and Propagation}, vol.~62,
  no.~6, pp. 3242--3256, 2014.

\bibitem{jaeckel2014quadriga2}
------, ``Quadriga: A 3-d multi-cell channel model with time evolution for
  enabling virtual field trials,'' \emph{IEEE Transactions on Antennas and
  Propagation}, vol.~62, no.~6, pp. 3242--3256, 2014.

\bibitem{qualcomm}
\BIBentryALTinterwordspacing
``{Deploying 5G NR mmWave to unleash the full 5G potential},'' QualComm, Tech.
  Rep., Nov. 2020. [Online]. Available:
  \url{https://www.qualcomm.com/content/dam/qcomm-martech/dm-assets/documents/deploying_mmwave_to_unleash_the_full_5g_potential_web.pdf}
\BIBentrySTDinterwordspacing

\bibitem{kingma2014adam}
J.~B. Diederik P.~Kingma, ``{ADAM}: A method for stochastic optimization,'' in
  \emph{Proceedings of International Conference on Learning Representations},
  2015.

\bibitem{vaswani2017attention}
A.~Vaswani, N.~Shazeer, N.~Parmar, J.~Uszkoreit, L.~Jones, A.~N. Gomez,
  {\L}.~Kaiser, and I.~Polosukhin, ``Attention is all you need,''
  \emph{Advances in Neural Information Processing Systems}, vol.~30, 2017.

\bibitem{bach2024learning}
F.~Bach, \emph{Learning Theory From First Principles}.\hskip 1em plus 0.5em
  minus 0.4em\relax Cambridge, MA, USA: MIT Press, To appear in 2024.

\end{thebibliography}

\end{document}